\documentclass[preprint,11pt]{elsarticle}

\usepackage{algorithmic}
\usepackage{algorithm}
\usepackage{natbib}
\usepackage{graphicx}
\usepackage{amsmath, amsthm}
\usepackage{vector, array, amssymb}
\usepackage{psfrag}
\usepackage{subfigure}
\usepackage{bbm}
\usepackage{enumitem}
\usepackage{color}
\usepackage{url}

\newtheorem{definition}{Definition}
\newtheorem{assumption}{Assumption}
\newtheorem{proposition}{Proposition}

\newtheorem{thm}{Theorem}
\newtheorem{cor}{Corollary}

\DeclareMathOperator{\Prob}{\text{Prob}}

\makeatletter
\newenvironment{proofof}[1]{\par
  \pushQED{\qed}%
  \normalfont \topsep6\p@\@plus6\p@\relax
  \trivlist
  \item[\hskip\labelsep
        \bfseries
    Proof of #1\@addpunct{.}]\ignorespaces
}{%
  \popQED\endtrivlist\@endpefalse
}
\makeatother

\begin{document}

\begin{frontmatter}

\title{On the connection between compression learning and scenario based optimization \tnoteref{t1}}
\tnotetext[t1]{Research was supported by the European Commission
under the projects MoVeS and SPEEDD. The authors would like to thank Prof. Simone Garatti for stimulating discussions and for bringing the work in reference \cite{Floyd_Warmuth1995} to our attention.}

\author[rvt]{Kostas ˜Margellos\corref{cor1}}
\ead{kostas.margellos@berkeley.edu}
\author[rvt1]{Maria Prandini}
\ead{prandini@elet.polimi.it}
\author[rvt2]{John ˜Lygeros}
\ead{lygeros@control.ee.ethz.ch}
\cortext[cor1]{Corresponding author}
\address[rvt]{Department of Industrial Engineering and Operations Research, UC Berkeley, Sutardja Dai Hall 330, Berkeley CA 94720, United States}
\address[rvt1]{Dipartimento di Elettronica, Informazione e Bioingegneria, Politecnico di Milano, Piazza Leonardo da Vinci 32, Milano 20133, Italy }
\address[rvt2]{Department of Information Technology and Electrical Engineering, ETH Z\"urich, Physikstrasse 3, Z\"urich 8092, Switzerland}

\begin{abstract}
We investigate the connections between compression learning and scenario based optimization.
We first show how to strengthen, or relax the consistency assumption at the basis of compression learning and study the learning and generalization properties of the algorithm involved.
We then consider different constrained optimization problems affected by uncertainty represented by means of scenarios. We show that the issue of providing guarantees on the probability of constraint violation reduces to a learning problem for an appropriately chosen algorithm that enjoys compression learning properties.
The compression learning perspective provides a unifying framework for scenario based optimization and allows us to
revisit the scenario approach and the probabilistically robust design, a recently developed technique based on a mixture of randomized and robust optimization, and to extend the guarantees on the probability of constraint violation to cascading optimization problems.
\end{abstract}


\end{frontmatter}

\section{Introduction} \label{sec: intro}
Optimal decision making in the presence of uncertainty is important for the efficient and economic operation of systems affected by endogenous, or exogenous uncertainties.
One approach to deal with uncertainty is through robust optimization. In this case a decision is made such that the constraints are satisfied for all admissible values of the uncertainty \cite{Robust_Opt_book}. Tractability of the developed techniques relies heavily on the geometry of the uncertainty set. On the other hand, chance constrained optimization allows for constraint violation but with an a-priori specified probability \cite{prekopa}, \cite{Shapiro}. In \cite{NemShap}, \cite{Berts2}, different approximations to chance constrained optimization problems are proposed for the case where the constraints exhibit a specific structure with respect to the uncertainty and under certain assumptions on the underlying probability distribution.

In many cases, however, we are only provided with data, e.g. historical values of the uncertainty. Therefore, research has been devoted towards the development of a data driven decision making paradigm. Under such a set-up, an alternative to robust optimization is scenario based optimization, that involves solving an optimization problem whose constraints depend only on a finite number of uncertainty instances called ``scenarios''. Scenario based optimization does not require any specific assumption on the probability distribution of the uncertainty or the way in which the uncertainty enters the optimization problem. On the other hand, it does require certain structure of the underlying optimization problem to ensure that the properties of the solution
generalize to unseen uncertainty instances and hence to provide guarantees regarding the probability of constraint satisfaction. For problems that are convex with respect to the decision variables the so called scenario approach \cite{Calafiore_Campi2006}, \cite{Campi_Garatti2008}, \cite{Calafiore}, offers an already mature theoretical framework for analyzing the generalization properties of the optimal solution. In the non-convex case, tools from statistical learning \cite{VCtheory}, \cite{Vapnik}, \cite{Anthony_Biggs} based on the VC theory can be employed to provide guarantees on the probability of constraint satisfaction for any feasible solution of an optimization problem \cite{Vidyasagar1997}, \cite{book_Tempo}, \cite{AlamoVC}.

In this paper we explore the links between learning theory and the scenario approach to scenario based optimization without resorting to VC theoretic results. To this end we exploit the results of \cite{Floyd_Warmuth1995} and consider compression learning algorithms, that are based on an alternative notion of learning under an assumption referred to as consistency. We first show
how using ideas from the scenario approach theory one can strengthen or relax the consistency assumption, and analyze the resulting learnability properties. We then return to optimization problems and show that the problem of providing guarantees regarding the probability of constraint violation can be equivalently thought of as a learning problem for an appropriately chosen algorithm that enjoys some compression property. Different classes of optimization programs from the literature are considered. In particular we revisit the scenario approach \cite{Calafiore_Campi2006}, \cite{Campi_Garatti2008}, \cite{Calafiore} and and the probabilistically robust design, a recently developed technique that is based on a mixture of randomized and robust optimization, proposed in \cite{Margellos2013}. Moreover, we consider the class of cascading optimization problems for which we provide novel results that offer guarantees regarding the probability of constraint satisfaction based on the compression learning perspective.

The rest of the paper unfolds as follows. Section \ref{sec: problem_statement} introduces the notion of compression. Section
\ref{sec: con_opt} shows how the learning theoretic results can be related to scenario based optimization and, in particular, the scenario approach and the probabilistically robust design. Section \ref{sec: seq_opt} shows how the proposed methodology can be employed for cascading optimization. Section \ref{sec: discussion} provides some discussion on the developed algorithms and elaborates on their relation with other learning based methodologies and Section \ref{sec: conclusion} provides some concluding remarks. To simplify the presentation of the paper the proofs of each section have been moved to the corresponding appendix.

\section{Learning results} \label{sec: problem_statement}
\subsection{Compression learning} \label{sec: conc_learning}
We start by describing some concepts and results from compression learning introduced in \cite{Floyd_Warmuth1995}.
We consider problems affected by an uncertain parameter $\delta$ taking values in some set $\Delta \subseteq \mathbb{R}^{n_\delta}$, which is endowed with a $\sigma$-algebra $\mathcal{D}$. Let $\mathbb{P}$ be a probability measure defined over $\mathcal{D}$. For $m \in \mathbb{N}$, we refer to a collection $\{\delta_i\}_{i=1}^{m}$ of $m$ i.i.d. samples $\delta_i \in \Delta$ extracted according to $\mathbb{P}$ as an $m$-multisample.
We will refer to the elements $\mathcal{D}$ as \emph{concepts}. For any concept $C \in \mathcal{D}$ let $\mathbbm{1}_C(\cdot): \Delta \rightarrow \{0,1\}$ be the standard indicator function of $C$, i.e. $\mathbbm{1}_C(\delta) = 1$ if $\delta \in C$ and zero otherwise.
Denote by $T \in \mathcal{D}$ a fixed but possibly unknown \emph{target concept} for which we assume that an oracle is available, that for any $\delta \in \Delta$, provides the labeling $\mathbbm{1}_T(\delta)$.
The following basic definitions are adapted from \cite{Vidyasagar1997}.

\begin{definition} \label{def: example}
[\emph{Labeled $m$-multisample}] Consider an $m$-multisample and a target concept $T \in \mathcal{D}$.
A labeled $m$-multisample is the collection $\big \{ \big ( \delta_i, \mathbbm{1}_T (\delta_i) \big ) \big \}_{i=1}^{m} \in [\Delta \times \{0,1\}]^m$.
\end{definition}

\begin{definition} \label{def: hypothesis}
[\emph{Consistent hypothesis}] Consider a labeled $m$-multisample and a target concept $T \in \mathcal{D}$. An element $H \in \mathcal{D}$ is called hypothesis.
$H$ is said to be consistent with the labeled $m$-multisample $\big \{ \big ( \delta_i, \mathbbm{1}_T (\delta_i) \big ) \big \}_{i=1}^{m}$ if and only if
$\mathbbm{1}_{H}(\delta_i) = \mathbbm{1}_{T}(\delta_i), \text{ for all } i=1,\ldots,m$.
\end{definition}
Definition \ref{def: hypothesis} implies that $H$ is a consistent hypothesis if it provides the same labeling of the samples $\delta_i$, $i=1, \dots, m$,  as  the target concept $T$. The error of $H$ as an approximation of the target concept $T$ can then be quantified through the probability measure of the set of uncertainty instances $\delta \in \Delta$ such that $H$ and $T$  give a different label. This error can be encoded by the measure of the symmetric difference of the sets $T$ and $H$, i.e.
\begin{align}
d_{\mathbb{P}}(T,H) = \mathbb{P} \big( \delta \in \Delta :~ \mathbbm{1}_{H} (\delta) \neq \mathbbm{1}_{T} (\delta) \big). \label{eq: error}
\end{align}
It is easy to see that $d_{\mathbb{P}}(T,H)$ is the measure\footnote{Throughout the paper we assume measurability of all involved sets. To relax this assumption the reader is referred to Appendix C in \cite{Grammatico_2013}.} of the symmetric difference of the sets $T$ and $H$. It is shown in \cite{Vidyasagar1997} that $d_{\mathbb{P}}(\cdot,\cdot)$ is not a metric, but just a pseudo-metric, since $d_{\mathbb{P}}(C_1,C_2) = 0$ does not imply that $C_1=C_2$, but only that the symmetric difference is a set of measure zero.

\begin{definition} \label{def: algorithm}
[\emph{Algorithm}] An algorithm is an indexed family of maps $\big \{ A_m \big \}_{m \geq m_0}$ for some $m_0 \in \mathbb{N}$. The map $A_m: [\Delta \times \{0,1\}]^m \rightarrow \mathcal{D}$ takes as input a labeled $m$-multisample and returns a hypothesis $A_m \big( \big \{ \big ( \delta_i, \mathbbm{1}_T (\delta_i) \big ) \big \}_{i=1}^{m} \big)$.
\end{definition}

The objective is to construct an approximation of the unknown target concept $T$ by constructing an algorithm such that the hypothesis $H_m = A_m \big( \big \{ \big ( \delta_i, \mathbbm{1}_T (\delta_i) \big ) \big \}_{i=1}^{m} \big)$ is consistent with the $m$-multisample.
Since $H_m$ depends on the extracted multisample, it is a random quantity defined on the product space $\Delta^m$ with measure $\mathbb{P}^m$. We can therefore state the quality of the obtained approximation only probabilistically, determining the probability with respect to $\mathbb{P}^m$ with which the approximation error $d_{\mathbb{P}}(T,H_m)$ exceeds a given threshold.

\begin{definition} \label{thm: pac}
[PAC-T algorithm] Let $T \in \mathcal{D}$ be a target concept. Suppose there exists $m_0 \in \mathbb{N}$ so that the algorithm $\big \{ A_m \big \}_{m \geq m_0}$ generates hypotheses $\big \{ H_m \big \}_{m \geq m_0}$ such that for any $\epsilon \in (0,1)$, $m \geq m_0$,

\begin{align}
\mathbb{P}^m \Big \{ (\delta_1,\ldots,\delta_m) \in \Delta^m :~ d_{\mathbb{P}}(T,H_m) > \epsilon \Big \} \leq q(m,\epsilon), \label{eq: pac}
\end{align}
for some function $q(m,\epsilon): \mathbb{N} \times (0,1) \rightarrow [0,1]$ such that $\lim_{m \rightarrow \infty} q(m,\epsilon) = 0$. Algorithm $\big \{ A_m \big \}_{m \geq m_0}$ is then said to be Probably Approximately Correct for the target concept $T$ (PAC-T).
\end{definition}

The statement of Definition \ref{thm: pac} is clearly related to PAC learnability \cite{Vidyasagar1997} (p. 56), where some concept class $\mathcal C \subseteq \mathcal D$ is considered and an algorithm is said to be PAC for the concept class $\mathcal{C}$ if \eqref{eq: pac} holds uniformly over target concepts $T \in \mathcal{C}$. Here we restrict attention to a specific target concept in view of the analysis of Section \ref{sec: con_opt}. For more details regarding PAC algorithms and PAC learnability the reader is referred to \cite{Vidyasagar1997}, \cite{Vapnik}.

Fix $d \in \mathbb{N}$ and consider $m \geq d$. We shall denote by $I_d = \{i_1,\ldots,i_d\} $ a set of $d$ indices from $\{1,\ldots,m\}$ and by $\mathcal{I}_d$ the set of cardinality ${m \choose d}$ containing all $I_d$ sets with $d$ indices.

 \begin{thm} \label{thm: compression}
[Thm. 5 in \cite{Floyd_Warmuth1995}] Let $T \in \mathcal{D}$ be a target concept. Fix $d \in \mathbb{N}$, consider $m>d$ and
denote by $G_d: [\Delta \times \{0,1\}]^d \rightarrow \mathcal{D}$ a map that, for any $I_d \in \mathcal{I}_d$, takes as input the labeled $d$-multisample $\big \{ \big ( \delta_i, \mathbbm{1}_T (\delta_i) \big ) \big \}_{i\in I_d}$ and returns a hypothesis\footnote{Unlike $H_m$, the subscript of $H_{I_d}$ is not an integer, but a set. The interpretation is that $H_{I_d}$ is the output of the compression function when fed with the samples $\{\delta_i\}_{i \in I_d}$. In the sequel we use a similar notation when defining $H_{m_d}$ for $m_d \in \mathcal{I}_d$. The reader is asked to excuse the slight abuse of the notation.} $H_{I_d} = G_d \big( \big \{ \big ( \delta_i, \mathbbm{1}_T (\delta_i) \big ) \big \}_{i \in I_d} \big)$ consistent with $\big \{ \big ( \delta_i, \mathbbm{1}_T (\delta_i) \big ) \big \}_{i\in I_d}$.
Then, for any $\epsilon \in (0,1)$ and any $m \geq d$
\begin{align}
\mathbb{P}^m \Big \{ &(\delta_1,\ldots,\delta_m) \in \Delta^m :~ \text{ there exists } I_d \in \mathcal{I}_d \text{ such that } \nonumber \\
& H_{I_d} \text{ is consistent with } \big \{ \big ( \delta_i, \mathbbm{1}_T (\delta_i) \big ) \big \}_{i=1}^{m} \text{ and } d_{\mathbb{P}}(T,H_{I_d}) > \epsilon \Big \} \leq {m \choose d} (1-\epsilon)^{m-d}. \label{eq: compression_d}
\end{align}
\end{thm}

Since for a fixed $d$, $\lim_{m \rightarrow \infty} {m \choose d} (1-\epsilon)^{m-d} = 0$, Theorem \ref{thm: compression} implies that for a sufficiently high number of samples $m$, the probability that there exists a subset $I_d$ with cardinality $d$ of the $m$ samples such that the hypothesis $H_{I_d}$ generated by $G_d$ is consistent with respect to all $m$ samples but the approximation error exceeds $\epsilon$ is low.
This theorem was stated in \cite{Floyd_Warmuth1995} in the context of sample compression, where the map $G_d$ is referred to as the compression function.

\begin{assumption} \label{ass: alg}
Let $T \in \mathcal{D}$ be a target concept. Assume that
there exists $d$ and $G_d: [\Delta \times \{0,1\}]^d \rightarrow \mathcal{D}$ taking as input a labeled $d$-multisample
such that: \\
1) For all $I_d \in \mathcal{I}_d$, $H_{I_d}$ is consistent with $\big \{ \big ( \delta_i, \mathbbm{1}_T (\delta_i) \big ) \big \}_{i\in I_d}$. \\
2) With $\mathbb{P}^m$-probability one, for any labeled $m$-multisample $\big \{ \big ( \delta_i, \mathbbm{1}_T (\delta_i) \big ) \big \}_{i=1}^{m}$ with $m \geq d$, there exists $I_d \in \mathcal{I}_d$ such that
the hypothesis $H_{I_d} = G_d \big( \big \{ \big ( \delta_i, \mathbbm{1}_T (\delta_i) \big ) \big \}_{i \in I_d} \big)$ is consistent with the labeled $m$-multisample.
\end{assumption}

Assumption \ref{ass: alg} implies that any sufficiently large $m$-multisample can be compressed, i.e. there exists a subset of this multisample with fixed cardinality $d$ which we can use to generate a hypothesis that is consistent with the entire $m$-multisample.
The assumption that for any $I_d \in \mathcal{I}_d$, the hypothesis $H_{I_d}$ is consistent with the $d$-multisample used to construct it, is trivially satisfied for the optimization problems considered in the next section.

Under Assumption \ref{ass: alg}, let the map $m_d: [\Delta \times \{0,1\}]^m \rightarrow \mathcal{I}_d$ return a set of $d$ indices such that $G_d \big( \big \{ \big ( \delta_i, \mathbbm{1}_T (\delta_i) \big ) \big \}_{i \in m_d} \big)$ is consistent with the entire $\big \{ \big ( \delta_i, \mathbbm{1}_T (\delta_i) \big ) \big \}_{i=1}^{m}$. Construct the algorithm $\big \{ A_m \big \}_{m \geq d}$, where $A_m: [\Delta \times \{0,1\}]^m \rightarrow \mathcal{D}$ takes as input a labeled $m$-multisample and returns a hypothesis
\begin{align}
H_m = A_m \big( \big \{ \big ( \delta_i, \mathbbm{1}_T (\delta_i) \big ) \big \}_{i=1}^m \big) = G_d \big( \big \{ \big ( \delta_i, \mathbbm{1}_T (\delta_i) \big ) \big \}_{i \in m_d} \big).
\end{align}
We then have the following theorem, which is stated in \cite{Floyd_Warmuth1995} without a proof.

\begin{thm} \label{thm: prop_viol1}
[Thm. 6 in \cite{Floyd_Warmuth1995}] Let $T \in \mathcal{D}$ be a target concept. Under Assumption \ref{ass: alg}, algorithm $\big \{ A_{m} \big \}_{m \geq d}$ is PAC-T with $q(m,\epsilon) = {m \choose d} (1-\epsilon)^{m-d}$.
\end{thm}

\subsection{Strengthening the consistency assumption} \label{sec: imprBound}
Extending now the results of \cite{Floyd_Warmuth1995} we first show how the bound in Theorem \ref{thm: prop_viol1} can be tightened by slightly strengthening Assumption \ref{ass: alg}.
\begin{assumption} \label{ass: unique_I}
Let $T \in \mathcal{D}$ be a target concept. Assume that
there exists $d$ and $G_d: [\Delta \times \{0,1\}]^d \rightarrow \mathcal{D}$ taking as input a labeled $d$-multisample
such that:\\
1) For all $I_d \in \mathcal{I}_d$, $H_{I_d}$ is consistent with $\big \{ \big ( \delta_i, \mathbbm{1}_T (\delta_i) \big ) \big \}_{i\in I_d}$. \\
2) With $\mathbb{P}^m$-probability one, for any labeled $m$-multisample $\big \{ \big ( \delta_i, \mathbbm{1}_T (\delta_i) \big ) \big \}_{i=1}^{m}$ with $m \geq d$, there exists a unique $I_d \in \mathcal{I}_d$ such that
the hypothesis $H_{I_d} = G_d \big( \big \{ \big ( \delta_i, \mathbbm{1}_T (\delta_i) \big ) \big \}_{i \in I_d} \big)$ is consistent with the labeled $m$-multisample.
\end{assumption}
The addition over Assumption \ref{ass: alg} is that the set $I_d \in \mathcal{I}_d$ for which the requirements of Assumption
\ref{ass: unique_I} are satisfied is unique. For all $I_d \in \mathcal{I}_d$ define $S_{I_d} = \big \{ (\delta_1,\ldots,\delta_m) \in \Delta^m :~ H_{I_d} \text{ is consistent with }\\ \big \{ \big ( \delta_i, \mathbbm{1}_T (\delta_i) \big ) \big \}_{i=1}^m \big \}$, where $H_{I_d} = G_d \big( \big \{ \big ( \delta_i, \mathbbm{1}_T (\delta_i) \big ) \big \}_{i \in I_d} \big)$. We then have the following propositions which are used in the proof of Theorem \ref{thm: prop_viol1_impr}.
\begin{proposition} \label{prop: partition}
Under Assumption \ref{ass: unique_I}, $\{S_{I_d}\}_{I_d \in \mathcal{I}_d}$ forms a partition of $\Delta^m$ up to a set of measure zero, i.e. $\mathbb{P}^m \big \{ (\delta_1,\ldots,\delta_m) \in \Delta^m :~ \Delta^m \setminus \cup_{I_d \in \mathcal{I}_d} S_{I_d} \big \} = 0$ and
$S_{I_d^1} \cap S_{I_d^2} = \emptyset$ for all $I_d^1, I_d^2 \in \mathcal{I}_d$ with $I_d^1 \neq I_d^2$.
\end{proposition}

\begin{proposition} \label{prop: error_P}
Let $T \in \mathcal{D}$ be a target concept. Under Assumption \ref{ass: unique_I}, for any $I_d \in \mathcal{I}_d$ we have that
\begin{align}
F(\alpha) = \mathbb{P}^d \big \{ \{\delta_i\}_{i \in I_d} \in \Delta^d :~ d_{\mathbb{P}}(T,H_{I_d}) \leq \alpha \big \} = \alpha^d, \label{eq: errorF}
\end{align}
where $F(\cdot)$ is the probability distribution of the error $d_{\mathbb{P}}(T,H_{I_d})$ and $\alpha \in [0,1]$.
\end{proposition}
The proof of Proposition \ref{prop: error_P} is similar to the first part of the proof of Theorem 1 in \cite{Campi_Garatti2008}. Define $m_d$, $\big \{ A_m \big \}_{m \geq d}$ as in Section \ref{sec: conc_learning} and note that, under Assumption \ref{ass: unique_I}, $m_d: [\Delta \times \{0,1\}]^m \rightarrow \mathcal{I}_d$ is uniquely defined in this case.

\begin{thm} \label{thm: prop_viol1_impr}
Let $T \in \mathcal{D}$ be a target concept. Under Assumption \ref{ass: unique_I}, algorithm $\big \{ A_{m} \big \}_{m \geq d}$ is PAC-T with $q(m,\epsilon) = \sum_{i=0}^{d-1} {m \choose i} \epsilon^i (1-\epsilon)^{m-i}$ and in particular, for any $\epsilon \in (0,1)$ and any $m \geq d$,
\begin{align}
\mathbb{P}^m \Big \{ (\delta_1,\ldots,\delta_m) \in \Delta^m &:~ d_{\mathbb{P}}(T,H_{m}) > \epsilon \Big \} = \sum_{i=0}^{d-1} {m \choose i} \epsilon^i (1-\epsilon)^{m-i}. \label{eq: pac_tight}
\end{align}
\end{thm}

Theorem \ref{thm: prop_viol1_impr} constitutes a tighter version of Theorem \ref{thm: prop_viol1} since \eqref{eq: pac_tight} holds with equality for problems that satisfy Assumption \ref{ass: unique_I}. Moreover, the bound in the right-hand side of \eqref{eq: pac_tight} is tighter compared to the one in Theorem \ref{thm: prop_viol1}.
The proof of Theorem \ref{thm: prop_viol1_impr} is similar to the second part of the proof of Theorem 1 in \cite{Campi_Garatti2008}.

\subsection{Relaxing the consistency assumption} \label{sec: sampl_discard}
Finally, we revisit Theorem \ref{thm: prop_viol1} and investigate relaxing Assumption \ref{ass: alg}. To this end fix $r,~d \in \mathbb{N}$ and consider $m \geq d+r$. Given a set $I_r \in \mathcal{I}_r$, let the set $\mathcal{I}_d^{m-r}$ with cardinality ${m-r \choose d}$ contain all sets $I_d$ with $d$ indices from $\{ 1,\ldots,m\} \setminus I_r$.

\begin{assumption} \label{ass: alg_relax}
Let $T \in \mathcal{D}$ be a target concept. Assume that
there exists $d$ and $G_{d}: [\Delta \times \{0,1\}]^d \rightarrow \mathcal{D}$ taking as input a labeled $d$-multisample
such that: \\
1) For all $I_r \in \mathcal{I}_r$ and $I_d \in \mathcal{I}_d^{m-r}$, $H_{I_d}$ is consistent with $\big \{ \big ( \delta_i, \mathbbm{1}_T (\delta_i) \big ) \big \}_{i\in I_d}$.\\
2) With $\mathbb{P}^m$-probability one, for any labeled $m$-multisample $\big \{ \big ( \delta_i, \mathbbm{1}_T (\delta_i) \big ) \big \}_{i=1}^{m}$ with $m \geq d + r$, for all $I_r \in \mathcal{I}_r$ there exists $I_d \in \mathcal{I}_d^{m-r}$ such that
the hypothesis $H_{I_d} = G_d \big( \big \{ \big ( \delta_i, \mathbbm{1}_T (\delta_i) \big ) \big \}_{i \in I_d} \big)$ is consistent with $\big \{ \big ( \delta_i, \mathbbm{1}_T (\delta_i) \big ) \big \}_{i \in \{1,\ldots,m\} \setminus I_r}$, \\
3) With $\mathbb{P}^m$-probability one, for any labeled $m$-multisample $\big \{ \big ( \delta_i, \mathbbm{1}_T (\delta_i) \big ) \big \}_{i=1}^{m}$ with $m \geq d + r$, there exists $I_r \in \mathcal{I}_r$ such that for any $I_d \in \mathcal{I}_d^{m-r}$ that satisfies the first part of the assumption, the hypothesis $H_{I_d} = G_d \big( \big \{ \big ( \delta_i, \mathbbm{1}_T (\delta_i) \big ) \big \}_{i \in I_d} \big)$ is not consistent with $\big \{ \big ( \delta_i, \mathbbm{1}_T (\delta_i) \big ) \big \}$, for all $i \in I_r$.
\end{assumption}

The difference with Assumption \ref{ass: alg} is that we now allow $H_{I_d} = G_d \big( \big \{ \big ( \delta_i, \mathbbm{1}_T (\delta_i) \big ) \big \}_{i \in I_d} \big)$ to be inconsistent with $r$ elements of the labeled $m$-multisample.
Suppose that Assumption \ref{ass: alg_relax} is satisfied and denote by $\bar{I}_r \in \mathcal{I}_r$ the set of indices such that the third part of the assumption  holds. Let $\bar{m}_d^r: [\Delta \times \{0,1\}]^m \rightarrow \mathcal{I}_d$ be the map that for each labeled $m$-multisample $\big \{ \big ( \delta_i, \mathbbm{1}_T (\delta_i) \big ) \big \}_{i=1}^{m}$ returns a set of $d$ indices for which the corresponding hypothesis $G_d \big( \big \{ \big ( \delta_i, \mathbbm{1}_T (\delta_i) \big ) \big \}_{i \in \bar{m}_d^r} \big)$ is consistent with $\big \{ \big ( \delta_i, \mathbbm{1}_T (\delta_i) \big ) \big \}_{i \in \{1,\ldots,m\} \setminus \bar I_r}$ and is not consistent with $\big ( \delta_i, \mathbbm{1}_T (\delta_i) \big )$, for all ${i \in \bar{I}_r}$. Construct the algorithm $\big \{ A_m \big \}_{m \geq d+r}$, where $A_m: [\Delta \times \{0,1\}]^m \rightarrow \mathcal{D}$ takes as input a labeled $m$-multisample and returns a hypothesis $H_m = A_m \big( \big \{ \big ( \delta_i, \mathbbm{1}_T (\delta_i) \big ) \big \}_{i=1}^m \big) = G_d \big( \big \{ \big ( \delta_i, \mathbbm{1}_T (\delta_i) \big ) \big \}_{i \in \bar{m}_d^r} \big)$.

\begin{thm} \label{thm: prop_viol1_relax}
Let $T \in \mathcal{D}$ be a target concept and fix $r \in \mathbb{N}$. Under Assumption \ref{ass: alg_relax}, algorithm $\big \{ A_m \big \}_{m \geq d+r}$ is PAC-T with $q(m,\epsilon) = {m \choose d} \sum_{i=0}^{r} {m-d \choose i} \epsilon^i (1-\epsilon)^{m-d-i}$, i.e.
\begin{align}
\mathbb{P}^{m} \Big \{ (\delta_1,\ldots,\delta_m) \in \Delta^{m} :~ d_{\mathbb{P}}(T,H_{m}) > \epsilon \Big \} \leq {m \choose d} \sum_{i=0}^{r} {m-d \choose i} \epsilon^i (1-\epsilon)^{m-d-i}. \label{eq: pac_m_relax}
\end{align}
\end{thm}
The proof of Theorem \ref{thm: prop_viol1_relax} is similar to the proof of Theorem 2.1 in \cite{Garatti}.
We can strengthen Assumption \ref{ass: alg_relax} by requiring the set $I_d \in \mathcal{I}_d^{m-r}$ that satisfies its requirements to be unique.
Consider now the following assumption, which is a relaxed version of Assumption \ref{ass: unique_I}.
\begin{assumption} \label{ass: unique_I_relax}
Consider the set-up of Assumption \ref{ass: alg_relax}. Assume also that the set $I_d \in \mathcal{I}_d^{m-r}$ that satisfies the requirements of Assumption \ref{ass: alg_relax} is unique.
\end{assumption}

Consider the algorithm $\big \{ A_m \big \}_{m \geq d+r}$, as constructed above Theorem \ref{thm: prop_viol1_relax}.
We then have the following theorem.

\begin{thm} \label{thm: prop_viol1_impr_relax1}
Let $T \in \mathcal{D}$ be a target concept and fix $r \in \mathbb{N}$. Under Assumption \ref{ass: unique_I_relax}, algorithm $\big \{ A_m \big \}_{m \geq d+r}$ is PAC-T with $q(m,\epsilon) = {r + d -1 \choose r} \sum_{i=0}^{r + d -1} {m \choose i} \epsilon^i (1-\epsilon)^{m-i}$, i.e.
\begin{align}
\mathbb{P}^{m} \Big \{ (\delta_1,\ldots,\delta_m) \in \Delta^{m} :~ d_{\mathbb{P}}(T,H_{m}) > \epsilon \Big \} \leq {r + d -1 \choose r} \sum_{i=0}^{r + d -1} {m \choose i} \epsilon^i (1-\epsilon)^{m-i}. \label{eq: pac_m_relax1}
\end{align}
\end{thm}
The proof of Theorem \ref{thm: prop_viol1_impr_relax1} follows the proof of Theorem 2.1 in \cite{Garatti}.
It constitutes a variant of Theorem \ref{thm: prop_viol1_impr}
when Assumption \ref{ass: unique_I} is relaxed to Assumption \ref{ass: unique_I_relax}.
However, in contrast to Theorem \ref{thm: prop_viol1_impr}, the bound in \eqref{eq: pac_m_relax1} is not tight, since \eqref{eq: thm_compr8_relax}, \eqref{eq: thm_compr9_relax} in the proof of Theorem \ref{thm: prop_viol1_relax} do not hold with equality.

\section{Connection to optimization} \label{sec: con_opt}
\subsection{Scenario based optimization as a learning problem}
Consider the robust optimization problem
\begin{align}
\mathcal{P}: & \min_{x \in \mathcal{X}} c^T x  \nonumber\\
& \text{ subject to: } g(x,\delta) \leq 0,\, \forall \delta \in \Delta, \label{eq: robust}
\end{align}
where $\mathcal{X} \subset \mathbb{R}^{n_x}$, $c \in \mathbb{R}^{n_x}$ and $g: \mathcal{X} \times \Delta \rightarrow \mathbb{R}$. As in Section \ref{sec: problem_statement} we assume that $\Delta$ is endowed with a $\sigma$-algebra and a probability measure $\mathbb{P}$. We consider here only one scalar-valued constraint function without loss of generality; in case of multiple constraint functions $g_j: \mathcal{X} \times \Delta \rightarrow \mathbb{R}$, $j=1,\ldots,n_c$, we can set $g(x,\delta) = \max_{j=1,\ldots,n_c} g_j(x,\delta)$. Moreover, considering a linear objective function is also without loss of generality; in case we seek to minimize a generic objective function, an epigraphic reformulation could be employed \cite{Calafiore_Campi2006}. Optimization programs in the form of $\mathcal{P}$ are generally difficult to solve when $\Delta$ is a continuous set.

To determine an (approximate) solution to \eqref{eq: robust}, an alternative optimization problem can be constructed, involving a multi-sample $\{\delta_i\}_{i=1}^{m} \in \Delta^m$ of finite size $m \in \mathbb{N}$, where the samples are extracted i.i.d according to $\mathbb{P}$.
\begin{align}
\mathcal{P}[\{\delta_i\}_{i=1}^{m}]: & \min_{x \in \mathcal{X}} c^T x \nonumber \\
& \text{ subject to: } g(x,\delta) \leq 0, \forall \delta \in
S \big ( \{\delta_i\}_{i=1}^m \big ), \label{eq: sampled}
\end{align}
where $S \big ( \{\delta_i\}_{i=1}^m \big ) \subseteq \Delta$ is a set that depends on the multisample; several choice of $S$ will be presented in the sequel, among them $S \big ( \{\delta_i\}_{i=1}^m \big ) = \{\delta_i\}_{i=1}^m$.

In the set-up of Section \ref{sec: problem_statement}, let $T=\Delta$ be the target concept, so that $\mathbbm{1}_T(\delta)=1$ for all $\delta \in \Delta$. Fix $d \in \mathbb{N}$ and consider $m \geq d$ and any map $x_d: \Delta^d \rightarrow \mathcal{X}$.
Define then a map
$G_d: [\Delta \times \{0,1\}]^d \rightarrow \mathcal{D}$ such that for any $I_d \in \mathcal{I}_d$, it returns a hypothesis $H_{I_d}$ constructed as
\begin{align}
H_{I_d} = G_d \big( \big \{ \big ( \delta_i, \mathbbm{1}_T (\delta_i) \big ) \big \}_{i \in I_d} \big) = \big \{ \delta \in \Delta:~ g(x_{d}( \{ \delta_i \}_{i \in I_d}),\delta) \leq 0 \big \}. \label{hyp_G_d}
\end{align}
Since $T = \Delta$, for any $I_d \in \mathcal{I}_d$, $d_{\mathbb{P}}(T,H_{I_d})$ is the probability of constraint violation, i.e.
\begin{align}
d_{\mathbb{P}}(T,H_{I_d}) = \mathbb{P}(\{\delta \in \Delta:~ \delta \notin H_{I_d}\}) = \mathbb{P} \big(\{ \delta \in \Delta:~ g(x_d(\{ \delta_i \big \}_{i \in I_d}),\delta) > 0 \}\big).
\end{align}

Suppose that $d$, $G_d$ are such that Assumption \ref{ass: alg} is satisfied.
Then there exists $m_d(\delta_1,\ldots,\delta_m) \in \mathcal{I}_d$ such that $H_{m_d} = \big \{ \delta \in \Delta :~ g(x_{d}(\{ \delta_i \big \}_{i \in m_d}),\delta) \leq 0 \big \}$ is consistent with $\big \{ \big ( \delta_i, \mathbbm{1}_T (\delta_i) \big ) \big \}_{i=1}^{m}$. Note that Assumption \ref{ass: alg} implicitly requires $H_{m_d}$ to be non-empty, since it must include $\{\delta_i\}_{i=1}^m$. This implies that $x_{d}(\{ \delta_i \big \}_{i \in m_d})$ is feasible for $\mathcal{P}[\{\delta_i\}_{i=1}^m]$.

\begin{thm} \label{thm: prop_viol}
Let $T = \Delta$ be the target concept and consider Assumption \ref{ass: alg}.
Let $x_m: \Delta^m \rightarrow \mathcal{X}$ be such that $x_m(\{ \delta_i \big \}_{i=1}^m) = x_{d}(\{ \delta_i \big \}_{i \in m_d})$ for a set $m_d \in \mathcal{I}_d$ that satisfies the second part of Assumption \ref{ass: alg}.
Then, for any $\epsilon \in (0,1)$,
\begin{align}
\mathbb{P}^m \Big \{ (\delta_1,\ldots,\delta_m) \in \Delta^m &:~ \mathbb{P} \big ( \delta \in \Delta :~ g(x_{m}(\{\delta_i\}_{i=1}^m),\delta) > 0 \big ) > \epsilon \Big \} \leq {m \choose d} (1-\epsilon)^{m-d}. \label{eq: prob_viol}
\end{align}
\end{thm}
Theorem \ref{thm: prop_viol} shows that under Assumption \ref{ass: alg}, for any feasible solution $x_m$ of $\mathcal{P}[\{\delta_i\}_{i=1}^{m}]$ such that $x_m(\{ \delta_i \big \}_{i=1}^m) = x_{d}(\{ \delta_i \big \}_{i \in m_d})$,
we can provide probabilistic guarantees regarding its feasibility of the form of \eqref{eq: prob_viol}. Note that the statement of \eqref{eq: prob_viol} shows that, with probability at least $1-{m \choose d} (1-\epsilon)^{m-d}$, $x_{m}$ satisfies \eqref{eq: robust} except for a set with $\mathbb{P}$-measure at most $\epsilon$.
The proof of Theorem \ref{thm: prop_viol} is based on showing that an algorithm is PAC-T for the target concept $T = \Delta$. This algorithm can be constructed as
$\big \{ A_m \big \}_{m \geq d}$, where $A_m: [\Delta \times \{0,1\}]^m \rightarrow \mathcal{D}$ is such that $H_m = A_m \big( \big \{ \big ( \delta_i, \mathbbm{1}_T (\delta_i) \big ) \big \}_{i=1}^m \big)$ and $H_m = H_{m_d}$. The hypothesis $H_m$ is defined as $H_{m} = \big \{ \delta \in \Delta:~ g(x_{m}( \{ \delta_i \}_{i=1}^m),\delta) \leq 0 \big \}$, while ensuring that $H_m = H_{m_d}$ is equivalent to $x_m(\{ \delta_i \big \}_{i=1}^m) = x_{d}(\{ \delta_i \big \}_{i \in m_d})$.
The latter is satisfied in the scenario approach set-up of Section \ref{sec: scenApp} and the probabilistically robust design of Section \ref{sec: boxApp}.

Note that if we replace Assumption \ref{ass: alg} with Assumption \ref{ass: unique_I}, Theorem \ref{thm: prop_viol} is still valid with the right-hand side of \eqref{eq: prob_viol} being replaced by the right-hand side of \eqref{eq: pac_tight} in Theorem \ref{thm: prop_viol1_impr}; in fact the result would hold with equality.
Following the discussion at the end of Section \ref{sec: sampl_discard}, one could also relax Assumption \ref{ass: alg} in a way such that the right-hand side of \eqref{eq: prob_viol} is replaced by
${r + d -1 \choose r} \sum_{i=0}^{r + d -1} {m \choose i} \epsilon^i (1-\epsilon)^{m-i}$. The interpretation of a hypothesis that is not consistent with some elements of the multi-sample in an optimization context is that we allow for some of the constraints to be violated. For problems that are convex with respect to the decision variables, this procedure is referred to as sampling-and-discarding in \cite{Garatti} and as constraint removal in \cite{Calafiore}.

We next consider problems for which probabilistic feasibility guarantees similar to \eqref{eq: prob_viol} are provided in \cite{Calafiore_Campi2006}, \cite{Margellos2013}, following a different methodology. Here we adopt the compression learning perspective and show that these problems share certain similarities, thus justifying the fact their guarantees are of the same form. In particular, we show that by appropriately selecting the constraint function, the uncertainty set of $\mathcal{P}[\{\delta_i\}_{i=1}^{m}]$ and the map $x_{m}: \Delta^m \rightarrow \mathcal{X}$, the requirements of Assumption \ref{ass: alg} are satisfied, and hence we obtain the probabilistic feasibility guarantees by virtue of Theorem \ref{thm: prop_viol}.

\subsection{The scenario approach} \label{sec: scenApp}
We first present the set-up of the scenario approach as this was proposed in \cite{Calafiore_Campi2006}. For any $m \in \mathbb{N}$ consider $\mathcal{P}[\{\delta_i\}_{i=1}^{m}]$ with $S(\{\delta_i\}_{i=1}^m)=\{\delta_i\}_{i=1}^m$; this results in the following optimization problem.
\begin{align}
\mathcal{P}_1[\{\delta_i\}_{i=1}^{m}]: & \min_{x \in \mathcal{X}} c^T x \nonumber \\
& \text{ subject to: } g(x,\delta) \leq 0, \forall \delta \in
\{\delta_i\}_{i=1}^m, \label{eq: scenApp}
\end{align}

Let $\mathcal{X}_m = \big \{ x \in \mathcal{X} :~ g(x,\delta) \leq 0, \forall \delta \in
\{\delta_i\}_{i=1}^m \big \}$ be the feasibility region of $\mathcal{P}_1[\{\delta_i\}_{i=1}^{m}]$ and consider the following assumption.
\begin{assumption} \label{ass: convex_scenApp}
The set $\mathcal{X} \subset \mathbb{R}^{n_x}$ is convex and for any $\delta \in \Delta$, the constraint function $g(\cdot,\delta)$ is convex.
For any $m$-multisample $\{\delta_i\}_{i=1}^m$, the feasibility region $\mathcal{X}_m$ of $\mathcal{P}_1[\{\delta_i\}_{i=1}^{m}]$ has a non-empty interior and the minimizer of $\mathcal{P}_1[\{\delta_i\}_{i=1}^{m}]$ exists and is unique.
\end{assumption}

The uniqueness and the feasibility part of the assumption can be relaxed as shown in \cite{Campi_Garatti2008}, \cite{Calafiore}. However, we keep these assumptions here to simplify the presentation.
Under Assumption \ref{ass: convex_scenApp}, let $x_m$ be the minimizer of $\mathcal{P}_1[\{\delta_i\}_{i=1}^{m}]$ and note that $x_m$ belongs to the feasibility region of $\mathcal{P}_1[\{\delta_i\}_{i=1}^{m}]$.

The scenario approach is based on the notion of support constraints. A constraint in $\mathcal{P}_1[\{\delta_i\}_{i=1}^{m}]$ is said to be a support constraint, if its removal results in an improvement in the objective value (see also Definition 4 in \cite{Calafiore_Campi2006}). In \cite{Calafiore}, under the convexity part of Assumption \ref{ass: convex_scenApp}, it is shown that, with $\mathbb{P}^m$-probability one,
the number of support constraints is bounded by the so called Helly's dimension. In \cite{Calafiore_Campi2006}, \cite{Campi_Garatti2008} it is shown that Helly's dimension is upper-bounded by $n_x$, whereas in \cite{Schildbach_2013} an improved bound is provided.
Let the number of support constraints be \emph{at most} $\zeta < \infty$.
Under Assumption \ref{ass: convex_scenApp}, and based on the definition of the support constraints, it can be shown that Assumption \ref{ass: alg} is satisfied for $d = \zeta$ and an appropriately constructed map $G_d$.

\begin{proposition} \label{prop: ass_scenApp}
Let $T = \Delta$ be the target concept and consider Assumption \ref{ass: convex_scenApp}. Fix $d = \zeta$ and consider $m \geq d$.
For any $I_d \in \mathcal{I}_d$, let $G_d: [\Delta \times \{0,1\}]^d \rightarrow \mathcal{D}$ return a hypothesis $H_{I_d} = \big \{ \delta \in \Delta:~ g(x_{d}( \{ \delta_i \}_{i \in I_d}),\delta) \leq 0 \big \}$, where $x_d$ is the minimizer of $\mathcal{P}_1[\{\delta_i\}_{i \in I_d}]$. $G_d$ then satisfies Assumption \ref{ass: alg}.
\end{proposition}
Under Proposition \ref{prop: ass_scenApp}, there exists $m_d \in \mathcal{I}_d$ with $d = \zeta$ such that the hypothesis $H_{m_d} = \big \{ \delta \in \Delta:~ g(x_{d}( \{ \delta_i \}_{i \in m_d}),\delta) \leq 0 \big \}$ is consistent with $\big \{ \big ( \delta_i, \mathbbm{1}_T (\delta_i) \big ) \big \}_{i =1}^m$.
Moreover, as shown in the proof of Proposition \ref{prop: ass_scenApp}, the set $m_d$ for which Assumption \ref{ass: alg} is satisfied is such that
$x_{d} \big ( \{ \delta_i \}_{i \in m_d} \big ) = x_m \big ( \{ \delta_i \}_{i =1 }^m \big )$, where $x_m$ is the unique (under Assumption \ref{ass: convex_scenApp}) minimizer of $\mathcal{P}_1[\{\delta_i\}_{i=1}^{m}]$.
This leads to the following corollary of Theorem \ref{thm: prop_viol}.

\begin{cor} \label{lm: lemma_scenApp}
Let $T = \Delta$ be the target concept and consider Assumption \ref{ass: convex_scenApp}. Fix $d = \zeta$ and consider $m \geq d$.
Then, for any $\epsilon \in (0,1)$,
\begin{align}
\mathbb{P}^m \Big \{ (\delta_1,\ldots,\delta_m) \in \Delta^m :~ \mathbb{P} \big ( \delta \in \Delta :~ g(x_m,\delta) > 0 \big ) > \epsilon \Big \} \leq {m \choose d} (1-\epsilon)^{m-d}, \label{eq: prob_viol_scenApp}
\end{align}
where $x_m$ is the minimizer of $\mathcal{P}_1[\{\delta_i\}_{i=1}^{m}]$.
\end{cor}

Corollary \ref{lm: lemma_scenApp} provides guarantees on the probability that the optimal solution of $\mathcal{P}_1[\{\delta_i\}_{i=1}^{m}]$ violates the constraints.
Note that this result is identical to Theorem 1 of \cite{Calafiore_Campi2006} (with $n_x$ in place of $\zeta$) but is not the same with the refined bound of Theorem 1 of \cite{Campi_Garatti2008}.
To obtain the same conclusion with Theorem 1 of \cite{Campi_Garatti2008} we focus first on problems in the form of $\mathcal{P}_1[\{\delta_i\}_{i=1}^{m}]$ such that, with $\mathbb{P}^m$-probability one, the number of support constraints is \emph{equal to} $\zeta$. In the particular case where $d =\zeta = n_x$, we have the class of fully supported problems \cite{Campi_Garatti2008}. Considering problems
where $\mathcal{P}_1[\{\delta_i\}_{i=1}^{m}]$ has exactly $\zeta$ support constraints with probability one, is a sufficient condition for Assumption \ref{ass: unique_I} to be satisfied. This is summarized in the following proposition.

\begin{proposition} \label{prop: fully_sup}
Let $T = \Delta$ be the target concept and consider Assumption \ref{ass: convex_scenApp}. Fix $d = \zeta$ and consider $m \geq d$.
For any $I_d \in \mathcal{I}_d$, let $G_d: [\Delta \times \{0,1\}]^d \rightarrow \mathcal{D}$ return a hypothesis $H_{I_d} = \big \{ \delta \in \Delta:~ g(x_{d}( \{ \delta_i \}_{i \in I_d}),\delta) \leq 0 \big \}$, where $x_d$ is the minimizer of $\mathcal{P}_1[\{\delta_i\}_{i \in I_d}]$.
If $\mathcal{P}_1[\{\delta_i\}_{i=1}^{m}]$ has exactly $\zeta$ support constraints with $\mathbb{P}^m$-probability one, then $G_d$ satisfies Assumption \ref{ass: unique_I}.
\end{proposition}

We then have the following corollary.
\begin{cor} \label{lm: lemma_scenApp_impr}
Let $T = \Delta$ be the target concept and consider Assumption \ref{ass: convex_scenApp}. Suppose that $\mathcal{P}_1[\{\delta_i\}_{i=1}^{m}]$ has exactly $\zeta$ support constraints with $\mathbb{P}^m$-probability one.
Fix $d = \zeta$ and consider $m \geq d$.
Then, for any $\epsilon \in (0,1)$,
\begin{align}
\mathbb{P}^m \Big \{ (\delta_1,\ldots,\delta_m) \in \Delta^m :~ \mathbb{P} \big ( \delta \in \Delta :~ g(x_m,\delta) > 0 \big ) > \epsilon \Big \} = \sum_{i=0}^{d-1} {m \choose i} \epsilon^i (1-\epsilon)^{m-i}, \label{eq: prob_viol_scenApp_impr}
\end{align}
where $x_m$ is the minimizer of $\mathcal{P}_1[\{\delta_i\}_{i=1}^{m}]$.
\end{cor}

If the problem does not have exactly $\zeta$ support constraints with $\mathbb{P}^m$-probability one, we can still obtain similar probabilistic guarantees following \cite{Calafiore}, \cite{Campi_Garatti2008}. Specifically, it is shown that if a problem is non-degenerate (see \cite{Calafiore} for a definition of non-degenerate problems) and has at most $\zeta$ support constraints, then by a procedure called regularization it can be transformed to a different problem with exactly $\zeta$ support constraints. We can then bound the probability in the left-hand side of \eqref{eq: prob_viol} by the probability of constraint violation for the regularized problem, which is equal to $\sum_{i=0}^{\zeta-1} {m \choose i} \epsilon^i (1-\epsilon)^{m-i}$.
In \cite{Campi_Garatti2008} it is shown that this is also the case even for degenerate problems that do not have exactly $\zeta$ support constraints.

We can replace Assumption \ref{ass: alg} in Proposition \ref{prop: ass_scenApp} and Assumption \ref{ass: unique_I}
in Proposition \ref{prop: fully_sup} by Assumption \ref{ass: alg_relax} and Assumption \ref{ass: unique_I_relax}, respectively. The right-hand side of \eqref{eq: prob_viol_scenApp} is then replaced by the right-hand side of \eqref{eq: pac_m_relax}. Similarly, the right-hand side of \eqref{eq: prob_viol_scenApp_impr} is replaced by the right-hand side of
\eqref{eq: pac_m_relax1}, but the result does not necessarily hold with equality.
However, note that Assumption \ref{ass: convex_scenApp} does not suffice to ensure that both parts of Assumption \ref{ass: alg_relax} (similarly for Assumption \ref{ass: unique_I_relax}) are satisfied; it only guarantees (via Proposition \ref{prop: ass_scenApp} with $m-r$ in place of $m$) that the requirement of the first part holds.
To ensure that requirement of the second part is also satisfied we equip the algorithm constructed in the proof of Propositions \ref{prop: ass_scenApp} and \ref{prop: fully_sup} by a procedure that removes $r$ samples such that the minimizer of the problem with the remaining $m-r$ samples violates all constraints that correspond to the removed samples. As an effect of this removal procedure the objective value is always decreasing every time a sample is removed.

Such a procedure is referred to as sampling-and-discarding in \cite{Garatti} and as scenario approach with constraint removal in \cite{Calafiore}.
Moreover, in \cite{Garatti}, \cite{Calafiore}, different methodologies to construct such a procedure are proposed and their complexity is discussed: an optimal constraint removal scheme, however, with a combinatorial complexity; a greedy approach where the $r$ constraints to be removed are eliminated on a sequential fashion; and an approach based on the Lagrange multipliers associated with the constraint functions.

\subsection{Probabilistically robust design} \label{sec: boxApp}
We now revisit the probabilistically robust design proposed in \cite{Margellos2013}. For any $m \in \mathbb{N}$ consider the following optimization problem:
\begin{align}
\widetilde{\mathcal{P}}_2[\{\delta_i\}_{i=1}^{m}]: & \min_{\underline{p}, \overline{p} \in \mathbb{R}^{n_\delta}} ||\overline{p} - \underline{p}||_1 \nonumber \\
& \text{ subject to: } \delta \in \big [ \underline{p},~ \overline{p} \big ],  \forall \delta \in
\{\delta_i\}_{i=1}^m, \label{eq: BApp}
\end{align}
where the inclusion in \eqref{eq: BApp} should be interpreted element-wise. Denote by $p_m = (\underline{p}_{m},~ \overline{p}_{m}) \in \mathbb{R}^{2n_\delta}$ the minimizer of $\widetilde{\mathcal{P}}_2[\{\delta_i\}_{i=1}^{m}]$, which depends on the multisample $\{\delta_i\}_{i=1}^m$. Let $B(p_m) \subset \Delta$ be a hyper-rectangle constructed by the cartesian product of the intervals in $[ \underline{p}_m,~ \overline{p}_m \big ]$. Clearly, $B(p_m)$ is the smallest axis-aligned hyper-rectangle that contains all samples $\{\delta_i\}_{i=1}^m$.
Consider now the following optimization problem:
\begin{align}
\mathcal{P}_2[\{\delta_i\}_{i=1}^{m}]: & \min_{x \in \mathcal{X}} c^T x \nonumber \\
& \text{ subject to: } g(x,\delta) \leq 0, \forall \delta \in B(p_m). \label{eq: boxApp}
\end{align}
Problem $\mathcal{P}_2[\{\delta_i\}_{i=1}^{m}]$ is a robust program and requires the constraints to be satisfied for all values of the uncertainty inside $B(p_m)$, which is constructed based on the optimal solution of $\widetilde{\mathcal{P}}_2[\{\delta_i\}_{i=1}^{m}]$.
Note that $\mathcal{P}_2[\{\delta_i\}_{i=1}^{m}]$ is of the same form with $\mathcal{P}[\{\delta_i\}_{i=1}^{m}]$ with $S(\{\delta_i\}_{i=1}^m)=B\big (p_m(\{\delta_i\}_{i=1}^m) \big )$.
For a detailed discussion regarding conditions under which $\mathcal{P}_2[\{\delta_i\}_{i=1}^{m}]$ is tractable, the reader is referred to \cite{Margellos2013}.

Let now $\mathcal{X}_m = \{x \in \mathcal{X} :~ g(x,\delta) \leq 0,\, \forall \delta \in B(p_m) \}$ be the feasibility region of $\mathcal{P}_2[\{\delta_i\}_{i=1}^{m}]$
and consider the following assumption.
\begin{assumption} \label{ass: boxApp}
For any $m$-multisample $\{\delta_i\}_{i=1}^m$, the feasibility region $\mathcal{X}_m$ of $\mathcal{P}_2[\{\delta_i\}_{i=1}^{m}]$ has a non-empty interior, and the minimizer of $\mathcal{P}_2[\{\delta_i\}_{i=1}^{m}]$ exists and is unique.
\end{assumption}
Under Assumption \ref{ass: boxApp}, let $x_m$ to be the minimizer of $\mathcal{P}_2[\{\delta_i\}_{i=1}^{m}]$. Note that for any $\big \{ \delta_i \big \}_{i=1}^m$, $x_m(\{ \delta_i \big \}_{i=1}^m) \in \mathcal{X}_m$. Imposing the uniqueness assumption and selecting $x_m$ to be the minimizer of $\mathcal{P}_2[\{\delta_i\}_{i=1}^{m}]$ is to simplify the presentation of our results and at the end of the section we remove the uniqueness part of the assumption and discuss alternative choices for the map $x_m$.

\begin{proposition} \label{prop: ass_boxApp}
Let $T = \Delta$ be the target concept and consider Assumption \ref{ass: boxApp}. Fix $d = 2 n_\delta$ and consider $m \geq d$.
For any $I_d \in \mathcal{I}_d$, let $G_d: [\Delta \times \{0,1\}]^d \rightarrow \mathcal{D}$ return a hypothesis $H_{I_d} = \big \{ \delta \in \Delta:~ g(x_{d}( \{ \delta_i \}_{i \in I_d}),\delta) \leq 0 \big \}$, where $x_d$ is the minimizer of $\mathcal{P}_2[\{\delta_i\}_{i \in I_d}]$. $G_d$ then satisfies Assumption \ref{ass: alg}.
\end{proposition}

Under Proposition \ref{prop: ass_scenApp}, there exists $m_d \in \mathcal{I}_d$ with $d = 2 n_\delta$ such that the hypothesis $H_{m_d} = \big \{ \delta \in \Delta:~ g(x_{d}( \{ \delta_i \}_{i \in m_d}),\delta) \leq 0 \big \}$ is consistent with $\big \{ \big ( \delta_i, \mathbbm{1}_T (\delta_i) \big ) \big \}_{i =1}^m$. In fact, as shown in the proof of Proposition \ref{prop: ass_boxApp}, there exists a unique set of indices $m_d \in \mathcal{I}_d$ satisfying Assumption \ref{ass: alg}. Moreover, this set is such that $B(p_{d}(\{\delta_i\}_{i \in m_d})) = B(p_{m}(\{\delta_i\}_{i = 1}^m))$, where $p_d$ is the minimizer of $\widetilde{\mathcal{P}}_2[\{\delta_i\}_{i \in m_d}]$. The latter implies that $\mathcal{X}_{m_d} = \mathcal{X}_m$, where $\mathcal{X}_{m_d}$ is the feasibility region of $\mathcal{P}_2[\{\delta_i\}_{i \in m_d}]$. Due to the uniqueness part of Assumption \ref{ass: boxApp} we then have that $x_{d}( \{ \delta_i \}_{i \in m_d}) = x_m( \{ \delta_i \}_{i=1}^m )$, where $x_m$ is the minimizer of $\mathcal{P}_2[\{\delta_i\}_{i=1}^{m}]$.
This leads to the following corollary of Theorem \ref{thm: prop_viol}.

\begin{cor} \label{lm: lemma_boxApp}
Let $T = \Delta$ be the target concept and consider Assumption \ref{ass: boxApp}. Fix $d = 2n_\delta$ and consider $m \geq d$.
Then, for any $\epsilon \in (0,1)$,
\begin{align}
\mathbb{P}^m \Big \{ (\delta_1,\ldots,\delta_m) \in \Delta^m :~ \mathbb{P} \big ( \delta \in \Delta :~ g(x_m,\delta) > 0 \big ) > \epsilon \Big \} \leq {m \choose d} (1-\epsilon)^{m-d}, \label{eq: prob_viol_boxApp}
\end{align}
where $x_m$ is the minimizer of $\mathcal{P}_2[\{\delta_i\}_{i=1}^{m}]$.
\end{cor}

In general we can provide guarantees in the form of \eqref{eq: prob_viol_boxApp} for any a-priori specified map $x_{m}$ that determines some feasible solution of $\mathcal{P}_2[\{\delta_i\}_{i =1}^{ m}]$, and not only for the minimizer. Remove now the uniqueness requirement of Assumption \ref{ass: boxApp}. We show that, by selecting $x_{d}$ according to the following procedure, we obtain guarantees for the entire feasibility region $\mathcal{X}_m$ of the robust problem $\mathcal{P}_2[\{\delta_i\}_{i=1}^{m}]$.
To achieve this, for any $I_d \in \mathcal{I}_d$, consider the worst case probability of constraint violation $\sup_{x \in \mathcal{X}_{I_d}} \mathbb{P} \big ( \delta \in \Delta :~ g(x,\delta) > 0 \big )$. Then, for any $\bar{\epsilon} >0$ there exists $x_d[\bar{\epsilon}] : \Delta^d \rightarrow \mathcal{X}$ with $x_{d}[\bar{\epsilon}](\{\delta_i\}_{i \in I_d}) \in \mathcal{X}_{I_d}$ such that
\begin{align}
\sup_{x \in \mathcal{X}_{I_d}} \mathbb{P} \big ( \delta \in \Delta :~ g(x,\delta) > 0 \big ) < \mathbb{P} \big ( \delta \in \Delta :~ g(x_{d}[\bar{\epsilon}],\delta) > 0 \big ) + \bar{\epsilon}. \label{eq: prob_feas_region}
\end{align}
For $\bar{\epsilon} > 0$ pick any such $x_{d}[\bar{\epsilon}]$. Under this choice it can be shown that \eqref{eq: prob_viol_boxApp} can be replaced by
\begin{align}
\mathbb{P}^m \Big \{ (\delta_1,\ldots,\delta_m) \in \Delta^m :~ \sup_{x \in \mathcal{X}_m}\mathbb{P} \big ( \delta \in \Delta :~ g(x,\delta) > 0 \big ) > \epsilon \Big \} \leq {m \choose d} (1-\epsilon)^{m-d}. \label{eq: prob_viol_boxApp_feas}
\end{align}
The proof of this statement is similar to the proof of Corollary \ref{lm: lemma_boxApp} and relies on the fact that for the set $m_d(\{\delta_i\}_{i=1}^m)$ of indices satisfying Assumption \ref{ass: alg}, for any $\bar{\epsilon} > 0$, $x_{d}[\bar{\epsilon}]$ satisfies \eqref{eq: prob_viol_boxApp} and, as shown in the proof of Proposition \ref{prop: ass_boxApp}, $\mathcal{X}_{m_d} = \mathcal{X}_m$. Equation \eqref{eq: prob_viol_boxApp_feas} follows then from \eqref{eq: prob_feas_region} and the fact that $\bar{\epsilon} > 0$ is arbitrary.

The result in \eqref{eq: prob_viol_boxApp_feas} is similar but not identical to Proposition 1 of \cite{Margellos2013} where a tighter bound is provided; however, we achieve these guarantees by means of Theorem \ref{thm: prop_viol} without resorting to the scenario approach as in \cite{Margellos2013}.
The rest of the section demonstrates how we can obtain the same conclusion with Proposition 1 of \cite{Margellos2013}. To this end consider the following proposition.
\begin{proposition} \label{prop: suff_box}
Let $T=\Delta$ be the target concept and consider Assumption \ref{ass: boxApp}. Fix $d = 2 n_\delta$ and consider $m \geq d$.
For any $I_d \in \mathcal{I}_d$, let $G_d: [\Delta \times \{0,1\}]^d \rightarrow \mathcal{D}$ return a hypothesis $H_{I_d} = \big \{ \delta \in \Delta:~ g(x_{d}( \{ \delta_i \}_{i \in I_d}),\delta) \leq 0 \big \}$, where $x_d$ is the minimizer of $\mathcal{P}_2[\{\delta_i\}_{i \in I_d}]$.
If, with $\mathbb{P}^m$-probability one, for any $I_d \in \mathcal{I}_d$
\begin{align}
\Big \{ \delta \in \Delta :~ g(x_{d}(\{\delta_i\}_{i \in I_d}),\delta) > 0 \Big \} = \Big \{ \delta \in \Delta :~ \delta \notin B(p_{d}(\{\delta_i\}_{i \in I_d})) \Big \}, \label{eq: suff_box}
\end{align}
then $G_d$ satisfies Assumption \ref{ass: unique_I}.
\end{proposition}
We then have the following corollary.

\begin{cor} \label{lm: lemma_boxApp_impr}
Let $T = \Delta$ be the target concept and consider Assumption \ref{ass: boxApp}. Fix $d = \zeta$ and consider $m \geq d$. Suppose that with $\mathbb{P}^m$-probability one,
equality \eqref{eq: suff_box} is also satisfied for any $I_d \in \mathcal{I}_d$.
Then, for any $\epsilon \in (0,1)$,
\begin{align}
\mathbb{P}^m \Big \{ (\delta_1,\ldots,\delta_m) \in \Delta^m :~ \mathbb{P} \big ( \delta \in \Delta :~ g(x_m,\delta) > 0 \big ) > \epsilon \Big \} = \sum_{i=0}^{d-1} {m \choose i} \epsilon^i (1-\epsilon)^{m-i}, \label{eq: prob_viol_boxApp_impr}
\end{align}
where $x_m$ is the minimizer of $\mathcal{P}_2[\{\delta_i\}_{i=1}^{m}]$.
\end{cor}

If \eqref{eq: suff_box} is not satisfied, Corollary \ref{lm: lemma_boxApp_impr} does not hold any more; this is not the case with Corollary \ref{lm: lemma_boxApp}.
However, by inspection of \eqref{eq: boxApp} we have that, for any $x \in \mathcal{X}_{m}$, if $\delta \in B(p_{m})$ then $g(x,\delta) \leq 0$. Since the last statement holds for any $x \in \mathcal{X}_{m}$ it will also hold for $x_{m}$.
Therefore, $\mathbb{P} \big ( \delta \in \Delta :~ g(x_{m},\delta) > 0 \big ) \leq \mathbb{P} \big ( \delta \in \Delta :~ \delta \notin B(p_{m}) \big )$, and hence $\mathbb{P}^m \big \{ (\delta_1,\ldots,\delta_m) \in \Delta^m ~: \mathbb{P} \big ( \delta \in \Delta :~ g(x_{m},\delta) > 0 \big ) > \epsilon \big \} \leq \mathbb{P}^m \big \{ (\delta_1,\ldots,\delta_m) \in \Delta^m ~: \mathbb{P} \big ( \delta \in \Delta :~ \delta \notin B(p_{m}) \big ) > \epsilon \big \}$.
The right-hand side of the previous inequality corresponds to the probability with respect to $\mathbb{P}^m$ that the probability of constraint violation of $\widetilde{\mathcal{P}}_2[\{\delta_i\}_{i =1}^m]$ exceeds $\epsilon$. The latter falls in the framework of the scenario approach and has by construction $\zeta = 2n_\delta$ support constraints. In fact, $\widetilde{\mathcal{P}}_2[\{\delta_i\}_{i = m}^m]$ is a fully-supported problem with $2n_\delta$ decision variables. Therefore, for any $\epsilon > 0$, Assumption \ref{ass: unique_I} is satisfied for this problem and Corollary \ref{lm: lemma_scenApp_impr} holds with $d =\zeta = 2n_\delta$. Hence,
\begin{align}
\mathbb{P}^m \Big \{ (\delta_1,\ldots,\delta_m) \in \Delta^m :~ \mathbb{P} \big ( \delta \in \Delta ~|~ \delta \notin B(p_{m}) \big ) > \epsilon \Big \} = \sum_{i=0}^{2n_\delta-1} {m \choose i} \epsilon^i \big (1-\epsilon)^{m-i}.
\end{align}
Therefore, in any case we have that
\begin{align}
\mathbb{P}^m \Big \{ (\delta_1,\ldots,\delta_m) \in \Delta^m :~ \mathbb{P} \big ( \delta \in \Delta :~ g(x_{m},\delta) > 0 \big ) > \epsilon \Big \} \leq \sum_{i=0}^{2n_\delta-1} {m \choose i} \epsilon^i (1-\epsilon)^{m-i}. \label{eq: prob_viol_boxApp_impr1}
\end{align}
Note, however, that \eqref{eq: prob_viol_boxApp_impr1} is not tight. Moreover, selecting the map $x_{d}$ as in \eqref{eq: prob_feas_region} we can provide guarantees for the entire feasibility region $\mathcal{X}_m$, and replace the left-hand side in \eqref{eq: prob_viol_boxApp_impr1} by $\mathbb{P}^m \Big \{ (\delta_1,\ldots,\delta_m) \in \Delta^m :~ \sup_{x \in \mathcal{X}_m} \mathbb{P} \big ( \delta \in \Delta :~ g(x,\delta) > 0 \big ) > \epsilon \Big \}$. However, due to \eqref{eq: prob_feas_region}, the inequality in \eqref{eq: prob_viol_boxApp_impr1} would be strict.

We can replace Assumption \ref{ass: alg} in Proposition \ref{prop: ass_boxApp} and Assumption \ref{ass: unique_I}
in Proposition \ref{prop: suff_box} to Assumption \ref{ass: alg_relax} and Assumption \ref{ass: unique_I_relax}, respectively. The right-hand side of \eqref{eq: prob_viol_boxApp} is then replaced by the right-hand side of \eqref{eq: pac_m_relax}. Similarly, the right-hand side of \eqref{eq: prob_viol_boxApp_impr} is replaced by the right-hand side of
\eqref{eq: pac_m_relax1}, but the result does not necessarily hold with equality.
However, note that Assumption \ref{ass: boxApp} does not suffice to ensure that both parts of Assumption \ref{ass: alg_relax} (similarly for Assumption \ref{ass: unique_I_relax}) are satisfied; it only guarantees (via Proposition \ref{prop: ass_scenApp} with $m-r$ in place of $m$) that the requirement of the first part holds. Similarly to the scenario approach set-up, to ensure that requirement of the second part is also satisfied we equip the algorithm constructed in the proof of Propositions \ref{prop: ass_boxApp} and \ref{prop: fully_sup} by a procedure that removes $r$ samples such that the minimizer of the problem with the remaining $m-r$ samples violates all constraints that correspond to the removed samples. As an effect of this removal procedure the objective value is always decreasing every time a sample is removed.

In \cite{Margellos2013} one such procedure is proposed. First $\mathcal{P}_2[\{\delta_i\}_{i=1}^m]$ is solved and the samples that correspond to the active constraints of $\widetilde{\mathcal{P}}_2[\{\delta_i\}_{i=1}^m]$ are identified. In fact these samples are the ones that lie on the facets of $B(p_m)$. From these samples remove $\delta_j$, for some $j \in \{1,\ldots,m\}$ that yields the highest reduction in the objective value of $\mathcal{P}_2[\{\delta_i\}_{i \in \{1,\ldots,m\} \setminus j }]$ (this implies that the feasibility region is enlarged). Typically, this step requires solving $2n_\delta$ (assuming no multiple samples on the same facet of $B(p_m)$) robust optimization problems. We then proceed the same way until $r$ samples are removed.
Similarly to the scenario approach, as an effect of this removal procedure the objective value of the robust problem is always decreasing every time a sample is removed.

Note that for $m \in \mathbb{N}$, we selected $B(p_m)$ to be a hyper-rectangle. However,
any other representation (e.g. sphere, polytope, ellipsoid) with fixed parametrization could have been chosen instead, by reformulating $\widetilde{\mathcal{P}}_2[\{\delta_i\}_{i=1}^{m}]$ as
a convex volume minimization problem. In that case our analysis would remain unchanged with $2 n_\delta$ being replaced by the dimension of the parametrization vector $p_m$. For example, if $B(p_m)$ is a sphere, we would need $n_\delta + 1$ parameters.

\section{Cascading optimization problems} \label{sec: seq_opt}
\subsection{Probabilistic performance guarantees}
We consider here the class of cascading optimization problems. Every problem in the cascade is a program that depends on uncertainty scenarios but also on the solution of the preceding problem, while the same uncertainty scenarios are used in all problems in the cascade. Such problems arise in different contexts (e.g. multi-objective optimization, bilinear descent type of algorithms, approximate dynamic programming), yet, to the best of our knowledge, obtaining guarantees regarding the probability of constraint violation for the solution comprising the solutions of the individual problems in the cascade has proven to be elusive. Our analysis provides such guarantees for a cascade of two problems, but our results can be immediately extended to the case of any finite number of cascading problems.

For any $m \in \mathbb{N}$, consider the following family of problems which is parametric in $x \in \mathcal{X}$:
\begin{align}
\widetilde{\mathcal{P}}[x,\{\delta_i\}_{i=1}^{m}]: & \min_{y \in \mathcal{Y}} \tilde{c}^T y \nonumber \\
& \text{ subject to: } \widetilde{g}(y,x,\delta) \leq 0, \forall \delta \in
\widetilde{S} \big ( \{\delta_i\}_{i=1}^m \big ), \label{eq: sampled_y}
\end{align}
where $x \in \mathcal{X}$ is the vector of decision variables of an optimization problem of the form of $\mathcal{P}[\{\delta_i\}_{i=1}^{m}]$ in \eqref{eq: sampled}, $\mathcal{Y} \subset \mathbb{R}^{n_y}$, $\tilde{c} \in \mathbb{R}^{n_y}$, $\widetilde{g}:~ \mathcal{Y} \times \mathcal{X} \times \Delta \rightarrow \mathbb{R}$, and $\widetilde{S} \big ( \{\delta_i\}_{i=1}^m \big ) \subseteq \Delta$.

Suppose that $\mathcal{P}[\{\delta_i\}_{i=1}^{m}]$ and $\widetilde{\mathcal{P}}[x,\{\delta_i\}_{i=1}^{m}]$, for all $x \in \mathcal{X}$, fall in the scenario approach set-up, i.e. $\widetilde{S} \big ( \{\delta_i\}_{i=1}^m \big ) = S \big ( \{\delta_i\}_{i=1}^m \big ) = \{\delta_i\}_{i=1}^m$. Therefore, we impose the following assumption.

\begin{assumption} \label{ass:_cascade}
Suppose that $\mathcal{P}[\{\delta_i\}_{i=1}^{m}]$ is in the form of $\mathcal{P}_1[\{\delta_i\}_{i=1}^{m}]$ in \eqref{eq: scenApp} satisfying Assumption \ref{ass: convex_scenApp}. Moreover, $\widetilde{S} \big ( \{\delta_i\}_{i=1}^m \big ) = \{\delta_i\}_{i=1}^m$, the set $\mathcal{Y} \subset \mathbb{R}^{n_y}$ is convex and for any $x \in \mathcal{X}$ and any $\delta \in \Delta$, the constraint function $\widetilde{g}(\cdot,x,\delta)$ is convex.
For any $x \in \mathcal{X}$ and any $m$-multisample $\{\delta_i\}_{i=1}^m$, the feasibility region $\big \{ y \in \mathcal{Y} :~ \widetilde{g}(y,x,\delta) \leq 0, \forall \delta \in
\{\delta_i\}_{i=1}^m \big \}$ of $\widetilde{\mathcal{P}}[x,\{\delta_i\}_{i=1}^{m}]$ has a non-empty interior and the minimizer of $\widetilde{\mathcal{P}}[x,\{\delta_i\}_{i=1}^{m}]$ exists and is unique.
\end{assumption}

We only need to invoke Assumption \ref{ass:_cascade} in the proof of Proposition \ref{prop: cons_cascade} and Theorem \ref{thm: cons_cascade_opt}, where a by-product of Proposition \ref{prop: ass_scenApp} is employed. Alternatively, we could assume that problems $\mathcal{P}[\{\delta_i\}_{i=1}^{m}]$ and $\widetilde{\mathcal{P}}[x,\{\delta_i\}_{i=1}^{m}]$, for all $x \in \mathcal{X}$, fall in the set-up of the probabilistically robust design and modify Assumption \ref{ass:_cascade} so that both problems satisfy Assumption \ref{ass: boxApp}.
Moreover, even if these problems belong to any problem class, the subsequent developments would still follow, as long as the solution of each problem does not alter if we only use the subset of the $m$-multisample returned by the compression function defined below.

Under Assumption \ref{ass:_cascade}, Proposition \ref{prop: ass_scenApp} implies that Assumption \ref{ass: alg} is satisfied for some $d_1 \in \mathbb{N}$, $G_{d_1}: [\Delta \times \{0,1\}]^{d_1} \rightarrow \mathcal{D}$, which for any $I_{d_1} \in \mathcal{I}_{d_1}$ returns $H_{I_{d_1}} = \big \{ \delta \in \Delta:~ g(x_{d_1}( \{ \delta_i \}_{i \in I_{d_1}}),\delta) \leq 0 \big \}$, where $x_{d_1}:~ \Delta^{d_1} \rightarrow \mathcal{X}$ is the minimizer of $\mathcal{P}[\{\delta_i\}_{i \in I_{d_1}}]$.
Similarly, for any $x \in \mathcal{X}$, Assumption \ref{ass: alg} is also satisfied for some $d_2 \leq n_y$, $\widetilde{G}_{d_2}[x]: [\Delta \times \{0,1\}]^{d_2} \rightarrow \mathcal{D}$, which for any $I_{d_2} \in \mathcal{I}_{d_2}$ returns the hypothesis $\widetilde{H}_{I_{d_2}}[x] = \widetilde{G}_{d_2}[x] \big( \big \{ \big ( \delta_i, \mathbbm{1}_T (\delta_i) \big ) \big \}_{i \in I_{d_2}} \big ) = \big \{ \delta \in \Delta:~ \widetilde{g}(y_{d_2}[x](\{ \delta_i \}_{i \in I_{d_2}}),x,\delta) \leq 0 \big \}$, where $y_{d_2}[x]: \Delta^{d_2} \rightarrow \mathcal{Y}$ is the unique, under Assumption \ref{ass:_cascade}, minimizer of $\widetilde{\mathcal{P}}[x,\{\delta_i\}_{i \in I_{d_2}}]$.

\begin{proposition} \label{prop: cons_cascade}
Let $T=\Delta$ be the target concept and consider Assumption \ref{ass: convex_scenApp}. Fix
$d = d_1 + d_2$ and consider $m \geq d$.
Construct $G_d^c: [\Delta \times \{0,1\}]^{d} \rightarrow \mathcal{D}$, such that for any $I_d \in \mathcal{I}_d$,
\begin{align}
G_d^c &\big( \big \{ \big ( \delta_i, \mathbbm{1}_T (\delta_i) \big ) \big \}_{i \in I_d} \big) = H_{I_{d}} \cap \widetilde{H}_{I_{d}}[x_d(\{\delta_i\}_{i \in I_d})] \label{eq:cascade_end}\\
& = \big \{ \delta \in \Delta :~ \big ( g(x_{d}(\{\delta_i\}_{i \in I_d}),\delta) \leq 0 \big ) \text{ and } \big ( \widetilde{g}(y_{d}[x_{d}(\{\delta_i\}_{i \in I_d})](\{\delta_i\}_{i \in I_d}),x_{d}(\{\delta_i\}_{i \in I_d}),\delta) \leq 0 \big ) \}. \nonumber
\end{align}
$G_d^c$ then satisfies Assumption \ref{ass: alg}.
\end{proposition}
Proposition \ref{prop: cons_cascade} shows that if there exist a compression function for two optimization problems, then there exists a compression function for the cascade of these problems, where the outcome of the latter depends on the solution of the former.
Under Proposition \ref{prop: cons_cascade}, there exists $m_d\big ( \{ \delta_i \}_{i=1}^m \big ) \in \mathcal{I}_d$ such that the hypothesis $H_{m_d}^c = G_d^c \big( \big \{ \big ( \delta_i, \mathbbm{1}_T (\delta_i) \big ) \big \}_{i \in m_d} \big)$ is consistent with $\big \{ \big ( \delta_i, \mathbbm{1}_T (\delta_i) \big ) \big \}_{i=1}^{m}$.

\begin{thm} \label{thm: cons_cascade_opt}
Let $T = \Delta$ be the target concept and consider Assumption \ref{ass:_cascade}.
Fix $d = d_1 + d_2$ and consider $m \geq d$.
Then, for any $\epsilon \in (0,1)$,
\begin{align}
&\mathbb{P}^m \Big \{ (\delta_1,\ldots,\delta_m) \in \Delta^m :~ \nonumber \\ &\mathbb{P} \Big ( \delta \in \Delta :~ \big ( g(x_{m},\delta) > 0 \big ) \text{ or } \big ( \widetilde{g}(y_{m}[x_{m}],x_{m},\delta) > 0 \big ) \Big ) > \epsilon \Big \} \leq {m \choose d} (1-\epsilon)^{m-d}, \label{eq: pac_cascade_opt}
\end{align}
where $x_m$ and $y_m[x_m]$ are the minimizers of $\mathcal{P}[\{\delta_i\}_{i=1}^{m}]$ and $\widetilde{\mathcal{P}}[x_m,\{\delta_i\}_{i=1}^m]$, respectively.
\end{thm}
Theorem \ref{thm: cons_cascade_opt} provides a bound on the probability with which $x_{m}$, $y_{m}$ violate either the constraints of $\mathcal{P}[\{\delta_i\}_{i=1}^{m}]$, or the constraints of $\widetilde{\mathcal{P}}[x_{m},\{\delta_i\}_{i=1}^{m}]$.
Its proof is based on showing that an algorithm, $\{A_m\}_{m \geq d}$, is PAC-T for the target concept $T = \Delta$. This algorithm comprises $A_m: [\Delta \times \{0,1\}]^m \rightarrow \mathcal{D}$ such that $H_m = A_m \big( \big \{ \big ( \delta_i, \mathbbm{1}_T (\delta_i) \big ) \big \}_{i=1}^m \big)$ and $H_m = H_{m_d}^c$. The hypothesis $H_m$ is defined as $H_{m} = \big \{ \delta \in \Delta :~ \big ( g(x_{m},\delta) \leq 0 \big ) \text{ or } \big ( \widetilde{g}(y_{m}[x_{m}],x_{m},\delta) \leq 0 \big ) \big \}$. Ensuring that $H_m = H_{m_d}$ is equivalent to $x_m(\{ \delta_i \big \}_{i=1}^m) = x_{d}(\{ \delta_i \big \}_{i \in m_d})$ and $y_m[x_m](\{ \delta_i \big \}_{i=1}^m) = y_d[x_d](\{ \delta_i \big \}_{i \in m_d})$. The latter follows from the proof of Proposition \ref{prop: ass_scenApp}.
We refer to $\big \{ A_m \big \}_{m \geq d}$ as cascading algorithm since it is constructed based on a cascade of two sequentially dependent hypotheses.

Note that, under Assumption \ref{ass:_cascade}, we need $\widetilde{\mathcal{P}}[x,\{\delta_i\}_{i=1}^{m}]$ to be feasible for any $x \in \mathcal{X}$. To relax this requirement consider the set
\begin{align}
F = \Big \{ (\delta_1,\ldots,\delta_m) \in \Delta^m :~ \forall x \in & \big \{ x \in \mathcal{X} :~ g(x,\delta) \leq 0, \forall \delta \in \{\delta_i\}_{i=1}^m \big \}, \nonumber \\
& \big \{ y \in \mathcal{Y} :~ \widetilde{g}(y,x,\delta) \leq 0, \forall \delta \in
\{\delta_i\}_{i=1}^m \big \} \neq \emptyset \Big \}. \label{eq: feas_setF}
\end{align}
$F$ is a restriction of $\Delta^m$ on the set of multisamples for which the second problem in the cascade has a non-empty feasibility region (feasibility of the first one is ensured under Assumption \ref{ass: convex_scenApp}), not for any $x \in \mathcal{X}$, but for any $x \in \big \{ x \in \mathcal{X} :~ g(x,\delta) \leq 0, \forall \delta \in \{\delta_i\}_{i=1}^m \big \}$.
The result of Theorem \ref{thm: cons_cascade_opt} will then still hold if we replace $\Delta^m$ with $F$ in \eqref{eq: pac_cascade_opt}.

Theorem \ref{thm: cons_cascade_opt} implies that the solution comprising the solutions of the individual problems in the cascade is feasible for the constraints of both problems. In certain cases one can obtain similar guarantees by formulating a single optimization problem that involves minimizing some convex objective function (e.g. the objective function of the last problem in the cascade) with respect to both $x \in \mathbb{R}^{n_x}$ and $y \in \mathbb{R}^{n_y}$, and subject to the constraints of both problems in the cascade. However, guarantees in the form of \eqref{eq: pac_cascade_opt} can be still provided only if the second problem in the cascade is jointly convex with respect to $x$ and $y$. This is not required with the proposed approach, and the second problem in the cascade is allowed to have an arbitrary dependence with respect to $x$ (see Assumption \ref{ass:_cascade}). Moreover, even if the constraint functions are convex with respect to the decision variables of both problems, solving a single program involving all constraints may result to solutions $x$, $y$ that are not optimal for the individual problems in the cascade, thus leading to a degraded objective value.
One example of a problem with constraint functions that are not jointly convex with respect to the decision variables $x$ and $y$ can be found in bilinear descent type of algorithms. Suppose we seek to minimize some convex objective function subject to constraints that should hold for all $\delta \in \{\delta_i\}_{i=1}^m$, and the constraint functions are bi-convex with respect to $x$ and $y$.
One way to deal with this problem is by applying an iterative procedure with an a-priori fixed number of iterations. We could arbitrarily fix $y = y_0$ and consider the problem of minimizing only with respect to $x$. The resulting problem would then be in the form of $\mathcal{P}[\{\delta_i\}_{i=1}^{m}]$. Let $x_m$ be the minimizer of this problem. We can then fix $x = x_m$ in the initial problem and minimize only with respect to $y$. If we do not follow such an iterative approach, since the problem is non-convex, to provide guarantees in the form of \eqref{eq: pac_cascade_opt} one should resort to VC theory, which involves, however, the computation of an upper bound of the VC dimension, which is not necessarily easy to determine.

Another important feature of the proposed approach is that in both $\mathcal{P}[\{\delta_i\}_{i=1}^{m}]$ and $\widetilde{\mathcal{P}}[x,\{\delta_i\}_{i=1}^{m}]$ the same samples $\{\delta_i\}_{i=1}^m$ are used. This is required, for example, in the stochastic model predictive control context considered in \cite{Deori_2013}, where a cascade of two scenario programs was formulated.
The first problem in the cascade was in the form of $\mathcal{P}[\{\delta_i\}_{i=1}^{m}]$ with the constraint function encoding the input constraints (depending on samples).
At the second problem in the cascade, the bound on the system sate was considered as a decision variable. The objective was to minimize this (soft) bound, subject to both input and state constraints (depending on the same samples with the first problem) and the additional constraint $c^T y \leq c^T x_m + \alpha$, where $x_m$ is the minimizer of the first problem, $y$ includes the decision variables of the second problem and $\alpha > 0$ is a pre-specified degradation parameter. The second problem is then also in the form of $\mathcal{P}[\{\delta_i\}_{i=1}^{m}]$. This two-step approach has a multi-objective nature since it allows us to relax the state constraints by deciding upon their bound in the second problem in the cascade, while ensuring that the objective value deteriorates at most by a fixed amount $\alpha$ compared to the value obtained at the first problem. In particular, the two problems in the cascade have the same decision variables, i.e. $x = y$ and $n_x = n_y$, and the set $F$ in \eqref{eq: feas_setF} is such that $F = \Delta^m$.
The same samples have to be employed in both problems, otherwise feasibility of the second problem is not guaranteed. This is also the case in bilinear descent type of algorithms, since by using the same samples at every problem in the cascade, the objective function is confined to decrease at every iteration of the algorithm.

Unfortunately, for cascading problems we cannot provide the tighter bound of Theorem \ref{thm: prop_viol1_impr} in Section \ref{sec: imprBound}. Even if we replace Assumption \ref{ass: alg} with Assumption \ref{ass: unique_I} in Proposition \ref{prop: cons_cascade}, there does not necessarily exists a \emph{unique} set $I_d \in \mathcal{I}_d$ with $d=d_1+d_2$ such that the map $G_d^c$, constructed as in Proposition \ref{prop: cons_cascade}, satisfies Assumption \ref{ass: unique_I} (see also the construction of a set $I_d$ that satisfies Assumption \ref{ass: alg} in the proof of Proposition \ref{prop: cons_cascade}).
However, one can relax Assumption \ref{ass: alg} in Theorem \ref{thm: cons_cascade_opt} to Assumption \ref{ass: alg_relax} and replace the right-hand side of \eqref{eq: pac_cascade_opt} according to Theorem \ref{thm: prop_viol1_relax}.
To ensure that the obtained solution violates the removed constraints, thus satisfying the second part of Assumption \ref{ass: alg_relax}, we can follow the sampling and discarding procedure outlined in \cite{Campi_Garatti2008}. Removing a sample according to this procedure results in a reduction in the objective value of the optimization problem involved.
In the cascading set-up, however, we have multiple objective functions and since both problems in the cascade are based on the same samples $\{\delta_i\}_{i=1}^m$, removing a sample affects the constraints in both problems. If for example we are interested, as in most applications, in the value of the last problem in the cascade, then
removing a sample does not necessarily lead to a reduction in that objective value, since it may result in a different solution of the first problem in the cascade, which in turn affects the solution of the second problem.
To incorporate this requirement in the removal procedure, we can eliminate a sample only if it results in a reduction in the objective value of the subproblem of interest.

\section{Discussion} \label{sec: discussion}
In Section \ref{sec: problem_statement} we showed that any algorithm that satisfies some consistency assumption (Assumption \ref{ass: alg} or some of its strengthened or relaxed versions) is PAC-T learnable. In other words, consistency is a sufficient condition for learnability of a fixed, but possibly unknown, target concept $T \in \mathcal{D}$. The results of Section \ref{sec: problem_statement} can be easily extended to eansure learnability of an entire concept class $\mathcal{C} \subseteq \mathcal{D}$, thus implying that the underlying algorithm is PAC in the sense of \cite{Vidyasagar1997}. However, following \cite{Vidyasagar1997}, \cite{Vapnik}, having a concept class with finite VC dimension (see \cite{Vidyasagar1997} for a concise definition), which is a measure of the ``richness'' of this class, is a sufficient condition for PAC learnability. Therefore, the analysis of Section \ref{sec: problem_statement} complements the standard learning theoretic results based on VC theory, since an algorithm that generates a consistent hypothesis can be PAC even if the underlying concept class has infinite VC dimension.

Note that we consider here a fixed, but possibly unknown probability measure $\mathbb{P}$. However, if we are interested in learning a concept class uniformly with respect to any measure in some given class, then finite VC dimension is both a sufficient and necessary condition for PAC learnability. In this case, any concept class for which a consistent algorithm exists, would also have finite VC dimension. The results of Section \ref{sec: problem_statement} can be then useful to provide tighter bounds without relying on the computation of the VC dimension, which might be a difficult task.

It should be also noted that Theorem \ref{thm: compression} has a VC theoretic counterpart. For concept classes with finite VC dimension this is known as the probability of one-sided constrained failure \cite{Anthony_Biggs}, \cite{Vidyasagar1997}, \cite{AlamoVC}, and for $m \geq 8 / \epsilon$ it is of the form

\begin{align}
\mathbb{P}^m \Big \{ &(\delta_1,\ldots,\delta_m) \in \Delta^m :~ \text{ there exists } I_{d_{VC}} \in \mathcal{I}_{d_{VC}} \text{ such that } \nonumber \\
& H_{I_{d_{VC}}} \text{ is consistent with } \big \{ \big ( \delta_i, \mathbbm{1}_T (\delta_i) \big ) \big \}_{i=1}^{m} \text{ and } d_{\mathbb{P}}(T,H_{I_{d_{VC}}}) > \epsilon \Big \} \leq 2 \sum_{i=0}^{d_{VC}} {2m \choose i} 2^{- \frac{\epsilon m }{2}}, \label{eq: compression_VC}
\end{align}
where $d_{VC}$ denotes the VC dimension. Despite the similarities between \eqref{eq: compression_VC} and \eqref{eq: compression_d}, the proofs of the corresponding statements are fundamentally different. However, it is shown in \cite{Floyd_Warmuth1995} that Assumption \ref{ass: alg} is satisfied with $d = d_{VC}$ for a specific concept class with finite VC dimension, namely the so called maximum class. Connections between \eqref{eq: compression_VC} and constraint violation properties of optimization problems can be found in \cite{AlamoVC}.

Following the analysis of Section \ref{sec: problem_statement} for a generic algorithm, in Section \ref{sec: con_opt} it was shown how the problem of providing guarantees regarding the probability of constraint satisfaction can be thought of as the problem of learning a specific target concept $T = \Delta$ for an algorithm that involves solving some optimization and generates a consistent hypothesis. Different examples were studied (the scenario approach, the probabilistically robust design, cascading optimization) to illustrate that these problems share certain similarities, thus justifying the reason that we obtain probabilistic performance guarantees of similar nature. In all cases, the probability that the measure of constraint violation exceeds a given threshold $\epsilon \in (0,1)$, is bounded by some function $q(m,\epsilon)$ such that $\lim_{m \rightarrow \infty} q(m,\epsilon) = 0$. The quantity $q(m,\epsilon)$ is the confidence with which we can provide constraint violation guarantees. In many applications it is of importance to compute explicit sample complexity bounds, i.e. determine the number of samples $m$ for which $q(m,\epsilon) \leq \beta$, for some confidence level $\beta \in (0,1)$. The reader is referred to \cite{Floyd_Warmuth1995}, \cite{Calafiore_Campi2006}, \cite{Calafiore}, for explicit bounds related to the involved $q(m,\epsilon)$ functions, and to \cite{AlamoVC}, \cite{Alamo} for further refinements.

\section{Concluding remarks} \label{sec: conclusion}
In this paper we considered a compression learning paradigm for algorithms that satisfy some consistency assumption. We first showed how one can strengthen or relax this assumption and analyzed the implications on the learnability properties. We then concentrated on scenario based optimization problems and showed that one can provide guarantees regarding the probability of constraint violation by treating them as learning problems. In this context, we also showed how novel probabilistic feasibility guarantees can be provided for cascading optimization problems. These novel results demonstrate how compressed learning can prove useful for scenario based multi-objective and sequential optimization problems.

\section*{Appendix A: Proofs of Section \ref{sec: problem_statement}}
\begin{proofof}{Theorem \ref{thm: compression}}
Consider any $\epsilon \in (0,1)$.
The left-hand side of \eqref{eq: compression_d} can be expressed as follows:
\begin{align*}
\mathbb{P}^m &\Big \{ \bigcup_{I_d \in \mathcal{I}_d} \big \{ (\delta_1,\ldots,\delta_m) \in \Delta^m:~  H_{I_d} \text{ is consistent with } \big \{ \big ( \delta_i, \mathbbm{1}_T (\delta_i) \big ) \big \}_{i=1}^{m} \text{ and } d_{\mathbb{P}}(T,H_{I_d}) > \epsilon \big \} \Big \}.
\end{align*}
From the subadditivity of $\mathbb{P}^m$ it then follows that
\begin{align}
& \mathbb{P}^m \Big \{ \bigcup_{I_d \in \mathcal{I}_d} \big \{ (\delta_1,\ldots,\delta_m) \in \Delta^m:~  H_{I_d} \text{ is consistent with } \big \{ \big ( \delta_i, \mathbbm{1}_T (\delta_i) \big ) \big \}_{i=1}^{m} \text{ and } d_{\mathbb{P}}(T,H_{I_d}) > \epsilon \big \} \Big \} \nonumber\\
& \leq \sum_{I_d \in \mathcal{I}_d} \mathbb{P}^m \Big \{ \big \{ (\delta_1,\ldots,\delta_m) \in \Delta^m:~  H_{I_d} \text{ is consistent with } \big \{ \big ( \delta_i, \mathbbm{1}_T (\delta_i) \big ) \big \}_{i=1}^{m} \text{ and } d_{\mathbb{P}}(T,H_{I_d}) > \epsilon \big \} \Big \}. \label{eq: thm_compr}
\end{align}

Without loss of generality fix $I_d=\{1,\dots,d\} \in \mathcal{I}_d$ and consider any $(\delta_1,\ldots,\delta_d)$ in the set $\bar{\Delta}^d = \{ (\delta_1,\ldots,\delta_d) \in \Delta^d :~ d_{\mathbb{P}}(T,H_{I_d}\big( T, \{\delta_i\}_{i=1}^{d} \big)) > \epsilon \}$.
Note that, under Assumption \ref{ass: alg}, $H_{I_d}$ is consistent with $\big \{ \big ( {\delta}_i, \mathbbm{1}_T ({\delta}_i) \big ) \big \}_{i \in I_d}$.
If $\bar{\Delta}^d$ is empty, then, the contribution of $I_d$ to the right-hand-side of \eqref{eq: thm_compr} becomes zero, otherwise, we have that
\begin{align}
\mathbb{P} \Big \{ \delta \in \Delta &:~  H_{I_d} \text{ is consistent with } \big ( \delta, \mathbbm{1}_T (\delta) \big ) \text{ and } d_{\mathbb{P}}(T,H_{I_d}) > \epsilon \Big \} \nonumber \\
&= \mathbb{P} \Big \{ \delta \in \Delta:~  H_{I_d} \text{ is consistent with } \big ( \delta, \mathbbm{1}_T (\delta) \big ) \Big \}
\nonumber \\
&= 1 - d_{\mathbb{P}}(T,H_{I_d}) \leq 1-\epsilon,
\end{align}
where the first step follows from the fact that $d_{\mathbb{P}}(T,H_{I_d}\big( T, \{\delta_i\}_{i=1}^{d} \big))$ does not depend on $\delta$ but only on $(\delta_1,\ldots,\delta_d) \in \bar{\Delta}^d$, and the second step follows from the definition of a consistent hypothesis (Definition \ref{def: hypothesis}).

Since the samples are extracted independently we have that
\begin{align}
\mathbb{P}^{m-d} &\Big \{ (\delta_{d+1},\ldots,\delta_m) \in \Delta^{m-d}:~  H_{I_d} \text{ is consistent with } \big \{ \big ( \delta_i, \mathbbm{1}_T (\delta_i) \big ) \big \}_{i=d+1}^{m} \text{ and } d_{\mathbb{P}}(T,H_{I_d}) > \epsilon \Big \} \nonumber \\
& = \prod_{j=d+1}^{m} \mathbb{P} \Big \{ \delta_j \in \Delta:~  H_{I_d} \text{ is consistent with } \big ( \delta_j, \mathbbm{1}_T (\delta_j) \big ) \text{ and } d_{\mathbb{P}}(T,H_{I_d}) > \epsilon \Big \} \nonumber \\
& \leq (1-\epsilon)^{m-d}. \label{eq: conditional_prob}
\end{align}
Since \eqref{eq: conditional_prob} holds for any $(\delta_1,\ldots,\delta_d) \in \bar\Delta^d$, we can rewrite the first quantity in \eqref{eq: conditional_prob} using the relevant conditional probability measure, denoted by $\Prob$. We then have
\begin{align}
\Prob \Big \{ \big \{ (\delta_{1},\ldots,\delta_m) \in \Delta^m:~  H_{I_d} &\text{ is consistent with } \big \{ \big ( \delta_i, \mathbbm{1}_T (\delta_i) \big ) \big \}_{i=1}^{m} \text{ and } d_{\mathbb{P}}(T,H_{I_d}) > \epsilon \big \} \nonumber \\
&\Big| \big \{ (\delta_{1},\ldots,\delta_d) \in \bar{\Delta}^d \big \} \Big \} \leq (1-\epsilon)^{m-d}.
\end{align}
Integrating with respect to the (conditional) probability of extracting a $d$-multisample $(\delta_{1},\ldots,\delta_d)$ from the set $\bar{\Delta}^d$ we get
\begin{align}
\mathbb{P}^m &\Big \{ (\delta_1,\ldots,\delta_m) \in \Delta^m:~  H_{I_d} \text{ is consistent with } \big \{ \big ( \delta_i, \mathbbm{1}_T (\delta_i) \big ) \big \}_{i=1}^{m} \text{ and } d_{\mathbb{P}}(T,H_{I_d}) > \epsilon \Big \} \nonumber\\
& \leq \int_{\Delta^d} (1-\epsilon)^{m-d} \mathbbm{1}_{\big \{ (\delta_{1},\ldots,\delta_d) \in \bar{\Delta}^d \big \}} \mathbb{P}^d \big ( \{\textrm{d}\delta_i\}_{i \in I_d} \big ) \leq (1-\epsilon)^{m-d},
\end{align}
for $I_d=\{1,\dots,d\} \in \mathcal{I}_d$. A similar reasoning can be applied to any  $I_d \in \mathcal{I}_d$, which by equation \eqref{eq: thm_compr} leads to:
\begin{align}
\mathbb{P}^m &\Big \{ \bigcup_{I_d \in \mathcal{I}_d} \big \{ (\delta_1,\ldots,\delta_m) \in \Delta^m:~  H_{I_d} \text{ is consistent with } \big \{ \big ( \delta_i, \mathbbm{1}_T (\delta_i) \big ) \big \}_{i=1}^{m} \text{ and } d_{\mathbb{P}}(T,H_{I_d}) > \epsilon \big \} \Big \} \nonumber\\
&\leq \sum_{I_d \in \mathcal{I}_d} \max \big \{ 0, (1-\epsilon)^{m-d} \big \} \leq {m \choose d} (1-\epsilon)^{m-d},
\end{align}
and concludes the proof.
\end{proofof}

\begin{proofof}{Theorem \ref{thm: prop_viol1}}
Consider any $\epsilon \in (0,1)$. Under Assumption \ref{ass: alg}, let $m_d( \{ \delta_i \}_{i=1}^m) \in \mathcal{I}_d$ be such that the hypothesis $H_{m_d} = G_d \big( \big \{ \big ( \delta_i, \mathbbm{1}_T (\delta_i) \big ) \big \}_{i \in m_d} \big)$ is consistent with $\big \{ \big ( \delta_i, \mathbbm{1}_T (\delta_i) \big ) \big \}_{i=1}^{m}$.
We then have that
\begin{align}
&\mathbb{P}^m \Big \{ (\delta_1,\ldots,\delta_m) \in \Delta^m:~ d_{\mathbb{P}}(T,H_{m_d}) > \epsilon \Big\} \nonumber \\
&= \mathbb{P}^m \Big \{ (\delta_1,\ldots,\delta_m) \in \Delta^m :~ d_{\mathbb{P}}(T,H_{m_d}) > \epsilon \Big\} \nonumber \\
&=\mathbb{P}^m \Big \{ (\delta_1,\ldots,\delta_m) \in \Delta^m :~  H_{m_d} \text{ is consistent with } \big \{ \big ( \delta_i, \mathbbm{1}_T (\delta_i) \big ) \big \}_{i=1}^{m} \nonumber \\
& ~~~~~~~~~~~~~~~~~~~~~~~~~~~~~~~~~~~~~~~~~~~~~~~~~~~~~~~~~~~~~\text{ and } d_{\mathbb{P}}(T,H_{m_d}) > \epsilon \Big \}, \label{eq: proof_Thm2}
\end{align}
where the last equality follows from Assumption \ref{ass: alg}.
Now since the last term is upper bounded by
\begin{align}
\mathbb{P}^m \Big \{ (\delta_1,\ldots,\delta_m) \in \Delta^m &:~ \text{ there exists } I_d \in \mathcal{I}_d \text{ such that } \nonumber \\
& H_{I_d} \text{ is consistent with } \big \{ \big ( \delta_i, \mathbbm{1}_T (\delta_i) \big ) \big \}_{i=1}^{m} \text{ and } d_{\mathbb{P}}(T,H_{I_d}) > \epsilon \Big \},
\end{align}
by Theorem \ref{thm: compression}, we have that
\begin{align}
\mathbb{P}^m \Big \{ (\delta_1,\ldots,\delta_m) \in \Delta^m :~ d_{\mathbb{P}}(T,H_{m_d}) > \epsilon \Big \} \leq {m \choose d} (1-\epsilon)^{m-d}.
\end{align}
Set $q(m,\epsilon) = {m \choose d} (1-\epsilon)^{m-d}$. Since ${m \choose d} \leq \Big( \frac{me}{d}\Big)^d$ (Lemma 4.3 of \cite{Vidyasagar1997}), we have that $\lim_{m \rightarrow \infty} q(m,\epsilon) \\ \leq  \lim_{m \rightarrow \infty}  \Big( \frac{me}{d}\Big)^d (1-\epsilon)^{m-d} = 0$. Therefore, $\lim_{m \rightarrow \infty} q(m,\epsilon) = 0$. Construct then algorithm $\big \{ A_m \big \}_{m \geq d}$, where $A_m: [\Delta \times \{0,1\}]^m \rightarrow \mathcal{D}$ takes as input a labeled $m$-multisample and returns a hypothesis $H_m = A_m \big( \big \{ \big ( \delta_i, \mathbbm{1}_T (\delta_i) \big ) \big \}_{i=1}^m \big)$ such that $H_m = H_{m_d}$. By Definition \ref{thm: pac}, algorithm $\big \{ A_m \big \}_{m \geq d}$ is PAC-T.
\end{proofof}

\begin{proofof}{Proposition \ref{prop: partition}}
We first show that $\mathbb{P}^m \big \{ (\delta_1,\ldots,\delta_m) \in \Delta^m :~ \Delta^m \setminus \cup_{I_d \in \mathcal{I}_d} S_{I_d} \big \} = 0$.
It is equivalent to show that $\cup_{I_d \in \mathcal{I}_d} S_{I_d} = \Delta^m$ up to a set of measure zero.
Clearly, $\cup_{I_d \in \mathcal{I}_d} S_{I_d} \subseteq \Delta^m$. Therefore, it suffices to show that $\cup_{I_d \in \mathcal{I}_d} S_{I_d} \supseteq \Delta^m$, i.e. if $(\delta_1,\ldots,\delta_m) \in \Delta^m$ then there exists $I_d \in \mathcal{I}_d$ such that $(\delta_1,\ldots,\delta_m) \in S_{I_d}$. With $\mathbb{P}^m$-probability one, the last statement follows from Assumption \ref{ass: unique_I} and the definition of $S_{I_d}$.

It remains to show that $S_{I_d^1} \cap S_{I_d^2} = \emptyset$ for all $I_d^1, I_d^2 \in \mathcal{I}_d$ with $I_d^1 \neq I_d^2$. For the sake of contradiction assume that there exist $I_d^1, I_d^2 \in \mathcal{I}_d$ with $I_d^1 \neq I_d^2$ such that $S_{I_d^1} \cap S_{I_d^2} \neq \emptyset$. By the definition of $S_{I_d^1}, S_{I_d^2}$, this implies that there exists $(\delta_1,\ldots,\delta_m) \in \Delta^m$ such that both hypotheses $H_{I_d^1}$ and $H_{I_d^2}$ are consistent with $\big \{ \big ( \delta_i, \mathbbm{1}_T (\delta_i) \big ) \big \}_{i=1}^m$. However, by Assumption \ref{ass: unique_I} for any $m$-multisample $\big \{ \big ( \delta_i, \mathbbm{1}_T (\delta_i) \big ) \big \}_{i=1}^m$ there exists a unique $I_d \in \mathcal{I}_d$ such that the corresponding hypothesis is consistent with respect to the $m$-multisample, establishing a contradiction.
\end{proofof}

\begin{proofof}{Proposition \ref{prop: error_P}}
Without loss of generality fix $I_d=\{1,\dots,d\} \in \mathcal{I}_d$ and consider any $(\delta_1,\ldots,\delta_d) \in \Delta^d$. Denote by $\alpha \big (\{\delta_i\}_{i \in I_d} \big) = d_{\mathbb{P}}(T,H_{I_d})$ the error between the hypothesis $H_{I_d}$ and the target concept $T$, and recall that, under the first part of Assumption \ref{ass: unique_I}, $H_{I_d}$ is consistent with $\big \{ \big ( \delta_i, \mathbbm{1}_T ({\delta}_i) \big ) \big \}_{i \in I_d}$.
We have that

\begin{align}
\mathbb{P} \Big \{ \delta \in \Delta &:~  H_{I_d} \text{ is consistent with } \big ( \delta, \mathbbm{1}_T (\delta) \big ) \Big \} \nonumber \\
&= 1 - d_{\mathbb{P}}(T,H_{I_d}) = 1-\alpha \big (\{{\delta}_i\}_{i \in I_d} \big).
\end{align}

Since the samples are extracted independently we have that
\begin{align}
\mathbb{P}^{m-d} &\Big \{ (\delta_{d+1},\ldots,\delta_m) \in \Delta^{m-d}:~  H_{I_d} \text{ is consistent with } \big \{ \big ( \delta_i, \mathbbm{1}_T (\delta_i) \big ) \big \}_{i=d+1}^{m} \Big \} \nonumber \\
& = \prod_{j=d+1}^{m} \mathbb{P} \Big \{ \delta_j \in \Delta:~  H_{I_d} \text{ is consistent with } \big ( \delta_j, \mathbbm{1}_T (\delta_j) \big ) \Big \} \nonumber \\
& = \Big(1-\alpha \big (\{{\delta}_i\}_{i \in I_d} \big) \Big)^{m-d}. \label{eq: conditional_prob_a}
\end{align}
Integrating over $\big \{(\delta_{1},\ldots,\delta_d) \in \Delta^d \big \}$ we get
\begin{align}
\mathbb{P}^m \Big \{ (\delta_1,\ldots,\delta_m) \in \Delta^m &:~  H_{I_d} \text{ is consistent with } \big \{ \big ( \delta_i, \mathbbm{1}_T (\delta_i) \big ) \big \}_{i=1}^{m} \Big \} \nonumber\\
& = \int_{\Delta^d} \Big(1-\alpha \big (\{{\delta}_i\}_{i \in I_d} \big) \Big)^{m-d} \mathbb{P}^d \big ( \{\textrm{d}\delta_i\}_{i \in I_d} \big ) \nonumber \\
& = \int_{0}^1 (1-\alpha)^{m-d} F(\textrm{d}\alpha), \label{eq: cond5}
\end{align}
where the last equality is due to a change of variables and $F(\alpha)$ is defined by \eqref{eq: errorF} and denotes the probability distribution of the error $d_{\mathbb{P}}(T,H_{I_d})$.

Since $S_{I_d} = \big \{ (\delta_1,\ldots,\delta_m) \in \Delta^m ~|~ H_{I_d} \text{ is consistent with } \big \{ \big ( \delta_i, \mathbbm{1}_T ({\delta}_i) \big ) \big \}_{i=1}^m \big \}$, \eqref{eq: cond5} implies that
\begin{align}
\mathbb{P}^m \Big \{ (\delta_1,\ldots,\delta_m) \in S_{I_d} \Big \} = \int_{0}^1 (1-\alpha)^{m-d} F(\textrm{d} \alpha). \label{eq: cond6}
\end{align}

Under Assumption \ref{ass: unique_I}, Proposition \ref{prop: partition} holds. Therefore, we have that $S_{I_d}$, $I_d \in \mathcal{I}_d$ form a partition of $\Delta^m$ up to a set of measure zero. Hence,
\begin{align}
\sum_{I_d \in \mathcal{I}_d} \mathbb{P}^m \Big \{ (\delta_1,\ldots,\delta_m) \in S_{I_d} \Big \} = 1. \label{eq: cond9}
\end{align}

No set $S_{I_d}$, $I_d \in \mathcal{I}_d$ is more likely than the others, therefore, $\mathbb{P}^m \Big \{ (\delta_1,\ldots,\delta_m) \in S_{I_d} \Big \}$ is the same for all $I_d \in \mathcal{I}_d$. This fact together with \eqref{eq: cond9} implies that ${m \choose d} \mathbb{P}^m \Big \{ (\delta_1,\ldots,\delta_m) \in S_{I_d} \Big \} = 1$. The last statement together with \eqref{eq: cond6} leads to
\begin{align}
{m \choose d} \int_{0}^1 (1-\alpha)^{m-d} F(\textrm{d}\alpha) = 1. \label{eq: cond10}
\end{align}
As shown in \cite{Campi_Garatti2008}, there is a unique $F(\cdot)$ that satisfies \eqref{eq: cond10}. Integration by parts shows that $F(\alpha) = \alpha^d$ satisfies \eqref{eq: cond10} and concludes the proof.
\end{proofof}

\begin{proofof}{Theorem \ref{thm: prop_viol1_impr}}
Consider any $\epsilon \in (0,1)$. Fix any $I_d \in \mathcal{I}_d$ and denote by $\alpha \big (\{\delta_i\}_{i \in I_d} \big) = d_{\mathbb{P}}(T,H_{I_d})$ the error between the hypothesis $H_{I_d}$ and the target concept $T$. We then have that
\begin{align}
\mathbb{P}^m &\Big \{ \bigcup_{I_d \in \mathcal{I}_d} \big \{ (\delta_1,\ldots,\delta_m) \in \Delta^m :~  H_{I_d} \text{ is consistent with } \big \{ \big ( \delta_i, \mathbbm{1}_T (\delta_i) \big ) \big \}_{i=1}^{m} \text{ and } d_{\mathbb{P}}(T,H_{I_d}) > \epsilon \big \} \Big \} \label{eq: thm_compr1_impr} \\
& = {m \choose d} \int_{\Delta^d} (1-\alpha\big (\{\delta_i\}_{i \in I_d} \big))^{m-d} \mathbbm{1}_{\big \{ \big (\{\delta_i\}_{i \in I_d} \big) \in \Delta^d :~ \alpha\big (\{\delta_i\}_{i \in I_d} \big) > \epsilon \big \}} \mathbb{P}^d \big ( \{\textrm{d}\delta_i\}_{i \in I_d} \big ) \label{eq: thm_compr2_impr} \\
& = {m \choose d} \int_{\epsilon}^1 (1-\alpha)^{m-d} F(\textrm{d}\alpha) \label{eq: thm_compr3_impr} \\
& = {m \choose d} \int_{\epsilon}^1 (1-\alpha)^{m-d} d \alpha^{d-1} \textrm{d}\alpha \label{eq: thm_compr4_impr} \\
& = \sum_{i=0}^{d-1} {m \choose i} \epsilon^i (1-\epsilon)^{m-i}, \label{eq: thm_compr5_impr}
\end{align}
where \eqref{eq: thm_compr2_impr} follows from \eqref{eq: thm_compr1_impr} using \eqref{eq: conditional_prob_a}, \eqref{eq: thm_compr3_impr} follows from \eqref{eq: thm_compr2_impr} by a change of variables and \eqref{eq: thm_compr4_impr} follows from the fact that, under Assumption \ref{ass: unique_I}, Proposition \ref{prop: error_P} implies that $F(\alpha) = \alpha^d$. Equality \eqref{eq: thm_compr5_impr} follows by repeated integration by parts (see also p. 1219 of \cite{Campi_Garatti2008}).

Under Assumption \ref{ass: unique_I}, let $m_d( \{ \delta_i \}_{i=1}^m) \in \mathcal{I}_d$ be the unique set of indices such that the hypothesis $H_{m_d} = G_d \big( \big \{ \big ( \delta_i, \mathbbm{1}_T (\delta_i) \big ) \big \}_{i \in m_d} \big)$ is consistent with $\big \{ \big ( \delta_i, \mathbbm{1}_T (\delta_i) \big ) \big \}_{i=1}^{m}$.
Since $m_d$ is unique, \eqref{eq: thm_compr1_impr} is equal to
\begin{align}
&\mathbb{P}^m \Big \{ (\delta_1,\ldots,\delta_m) \in \Delta^m :~  H_{m_d} \text{ is consistent with } \big \{ \big ( \delta_i, \mathbbm{1}_T (\delta_i) \big ) \big \}_{i=1}^{m} \nonumber \\
& ~~~~~~~~~~~~~~~~~~~~~~~~~~~~~~~~~~~~~~~~~~~~~~~~~~~~~~~~~~~~~\text{ and } d_{\mathbb{P}}(T,H_{m_d}) > \epsilon \Big \},
\end{align}
which based on \eqref{eq: proof_Thm2} (see proof of Theorem \ref{thm: prop_viol1}) is equal to $\mathbb{P}^m \Big \{ (\delta_1,\ldots,\delta_m) \in \Delta^m :~ d_{\mathbb{P}}(T,H_{m_d}) > \epsilon \Big \}$.
Therefore,
\begin{align}
\mathbb{P}^m \Big \{ (\delta_1,\ldots,\delta_m) \in \Delta^m :~ d_{\mathbb{P}}(T,H_{m_d}) > \epsilon \Big \} = \sum_{i=0}^{d-1} {m \choose i} \epsilon^i (1-\epsilon)^{m-i}. \label{eq: pac_impr}
\end{align}

Set $q(m,\epsilon) = \sum_{i=0}^{d-1} {m \choose i} \epsilon^i (1-\epsilon)^{m-i}$.
As shown in Lemma 4.3 of \cite{Vidyasagar1997}, ${m \choose i} \leq \Big( \frac{me}{i}\Big)^i$ for any $i \in \mathbb{Z}$. Therefore,
$q(m,\epsilon) = \sum_{i=0}^{d-1} \epsilon^i {m \choose i} (1-\epsilon)^{m-i} \leq \sum_{i=0}^{d-1} \epsilon^i \Big( \frac{me}{i}\Big)^i (1-\epsilon)^{m-i}$. Following the proof of Theorem \ref{thm: prop_viol1}, every term in last summation is such that $\lim_{m \rightarrow \infty} \Big( \frac{me}{i}\Big)^i (1-\epsilon)^{m-i} = 0$.  Therefore, $\lim_{m \rightarrow \infty} q(m,\epsilon) = 0$.
Construct then algorithm $\big \{ A_m \big \}_{m \geq d}$, where $A_m: [\Delta \times \{0,1\}]^m \rightarrow \mathcal{D}$ takes as input a labeled $m$-multisample and returns a hypothesis $H_m = A_m \big( \big \{ \big ( \delta_i, \mathbbm{1}_T (\delta_i) \big ) \big \}_{i=1}^m \big)$ such that $H_m = H_{m_d}$. By Definition \ref{thm: pac}, algorithm $\big \{ A_m \big \}_{m \geq d}$ is PAC-T.
\end{proofof}

\begin{proofof}{Theorem \ref{thm: prop_viol1_relax}}
Fix any $r \in \mathbb{N}$ and $I_r \in \mathcal{I}_r$, and under the second part of Assumption \ref{ass: alg_relax},
let $m_d^r \Big( \big \{ \big ( \delta_i, \mathbbm{1}_T (\delta_i) \big ) \big \}_{i \in \{1,\ldots,m\} \setminus I_r} \Big)$ be a set of $d$ indices such that $H_{m_d^r} = G_d \big( \big \{ \big ( \delta_i, \mathbbm{1}_T (\delta_i) \big ) \big \}_{i \in m_d^r} \big)$ is consistent with $\big \{ \big ( \delta_i, \mathbbm{1}_T (\delta_i) \big ) \big \}_{i \in \{1,\ldots,m\} \setminus I_r}$.
Denote then by $\alpha \big (\{\delta_i\}_{i \in \{1,\ldots,m\} \setminus I_r} \big) = d_{\mathbb{P}}(T,H_{m_d^r})$ the error between the hypothesis $H_{m_d^r}$ and the target concept $T$.
We then have that
\begin{align}
\mathbb{P} \Big \{ \delta \in \Delta &:~  H_{m_d^r} \text{ is not consistent with } \big ( \delta, \mathbbm{1}_T (\delta) \big ) \Big \} \nonumber \\
&= d_{\mathbb{P}}(T,H_{m_d^r}) = \alpha \big (\{\delta_i\}_{i \in \{1,\ldots,m\} \setminus I_r} \big).
\end{align}

Since the samples are extracted independently we have that
\begin{align}
\mathbb{P}^{r} &\Big \{ \{\delta_j\}_{j \in I_r} \in \Delta^{r}:~  H_{m_d^r} \text{ is not consistent with } \big \{ \big ( \delta_j, \mathbbm{1}_T (\delta_j) \big ) \big \}_{j \in I_r} \Big \} \nonumber \\
& = \prod_{j \in I_r} \mathbb{P} \Big \{ \delta_j \in \Delta:~  H_{m_d^r} \text{ is not consistent with } \big ( \delta_j, \mathbbm{1}_T (\delta_j) \big ) \Big \} \nonumber \\
& = \alpha \big (\{\delta_i\}_{i \in \{1,\ldots,m\} \setminus I_r} \big)^{r}. \label{eq: conditional_prob_a_relax}
\end{align}
Consider any $\epsilon \in (0,1)$. We then have that
\begin{align}
\mathbb{P}^m &\Big \{ (\delta_1,\ldots,\delta_m) \in \Delta^m :~  H_{m_d^r} \text{ is not consistent with } \big \{ \big ( \delta_i, \mathbbm{1}_T (\delta_i) \big ) \big \}_{i \in I_r} \text{ and } d_{\mathbb{P}}(T,H_{m_d^r}) > \epsilon \Big \} \label{eq: thm_compr1_relax} \\
& = \int_{\Delta^{m-r}} \alpha \big (\{\delta_i\}_{i \in \{1,\ldots,m\} \setminus I_r} \big)^{r} \mathbbm{1}_{\big \{ \alpha \big (\{\delta_i\}_{i \in \{1,\ldots,m\} \setminus I_r} \big) > \epsilon \big \}} \mathbb{P}^{m-r} \big ( \{\textrm{d}\delta_i\}_{i \in \{1,\ldots,m\} \setminus I_r} \big ) \label{eq: thm_compr2_relax} \\
& = \int_{\epsilon}^1 \alpha^{r} \bar{F}(\textrm{d}\alpha), \label{eq: thm_compr3_relax}
\end{align}
where \eqref{eq: thm_compr2_relax} follows from \eqref{eq: thm_compr1_relax} using \eqref{eq: conditional_prob_a_relax}, \eqref{eq: thm_compr3_relax} follows from \eqref{eq: thm_compr2_relax} by a change of variables and $\bar{F}(\cdot)$ is the probability distribution of the error $d_{\mathbb{P}}(T,H_{m_d^r})$, i.e.
\begin{align}
\bar{F}(\alpha) = \mathbb{P}^{m-r} \big \{ \{\delta_i\}_{i \in \{1,\ldots,m\} \setminus I_r} \in \Delta^{m-r} :~ d_{\mathbb{P}}(T,H_{m_d^r}) \leq \alpha \big \}. \label{eq: errorF_bar}
\end{align}
Notice the difference between \eqref{eq: errorF_bar} and \eqref{eq: errorF}; the latter is the distribution of the error between the target concept and the hypothesis generated using all elements of the multisample, whereas the former is the distribution of the error between the target concept and the hypothesis using only $d$ out of the $m-r$ elements of the multisample.

Construct the algorithm $\big \{ A_{m-r} \big \}_{m-r \geq d}$, where $A_{m-r}: [\Delta \times \{0,1\}]^{m-r} \rightarrow \mathcal{D}$ takes as input a labeled $m-r$-multisample and returns a hypothesis $H_{m-r} = A_{m-r} \big( \big \{ \big ( \delta_i, \mathbbm{1}_T (\delta_i) \big ) \big \}_{i \in \{1,\ldots,m\} \setminus I_r} \big)$ such that $H_{m-r} = H_{m_d^r}$.
By Theorem \ref{thm: prop_viol1} with $m-r$ in place of $m$ and $\alpha$ in place of $\epsilon$, we have that the constructed algorithm is PAC-T, hence
\begin{align}
\mathbb{P}^{m-r} \Big \{ \{\delta_i\}_{i \in \{1,\ldots,m\} \setminus I_r} \in \Delta^{m-r} :~ d_{\mathbb{P}}(T,H_{m_d^r}) > \alpha \Big \} \leq {m-r \choose d} (1-\alpha)^{m-r-d}. \label{eq: pac_m_r}
\end{align}
By \eqref{eq: errorF_bar}, \eqref{eq: pac_m_r}, we then have that $\bar{F}(\alpha) \geq 1 - {m-r \choose d} (1-\alpha)^{m-r-d}$ for any $\alpha \in [0,1]$.
The last statement together with \eqref{eq: thm_compr1_relax}-\eqref{eq: thm_compr3_relax} leads to

\begin{align}
\mathbb{P}^m &\Big \{ (\delta_1,\ldots,\delta_m) \in \Delta^m :~  H_{m_d^r} \text{ is not consistent with } \big \{ \big ( \delta_i, \mathbbm{1}_T (\delta_i) \big ) \big \}_{i \in I_r} \text{ and } d_{\mathbb{P}}(T,\bar{H}_{m-r}) > \epsilon \Big \} \nonumber \\
& = \int_{\epsilon}^1 \alpha^{r} \bar{F} (\textrm{d}\alpha) \label{eq: thm_compr0_relax} \\
& \leq \int_{\epsilon}^1 (m-r-d) {m-r \choose d} \alpha^{r} \big ( 1 - \alpha \big )^{m-r-d-1} \textrm{d}\alpha \label{eq: thm_compr4_relax} \\
& = {m-r \choose d} \frac{1}{ {m-d \choose r} } \sum_{i=0}^r {m-d \choose i} \epsilon^i (1-\epsilon)^{m-d-i},  \label{eq: thm_compr5_relax}
\end{align}
where the inequality in \eqref{eq: thm_compr4_relax} follows from \eqref{eq: thm_compr0_relax} due to the fact that $\bar{F}(\alpha) \geq 1 - {m-r \choose d} (1-\alpha)^{m-r-d}$ and is based on standard integral arguments (see also the proof of Theorem 2.1 in \cite{Garatti}).
Moreover, \eqref{eq: thm_compr5_relax} follows from \eqref{eq: thm_compr4_relax} by repeated integration by parts.

Denote now by $\bar{I}_r \in \mathcal{I}_r$ the set of indices for which the third part of Assumption \ref{ass: alg_relax} is satisfied. Let then $\bar{m}_d^r \Big( \big \{ \big ( \delta_i, \mathbbm{1}_T (\delta_i) \big ) \big \}_{i \in \{1,\ldots,m\} \setminus \bar{I}_r} \Big)$ be a set of $d$ indices such that the hypothesis $H_{\bar{m}_d^r} = G_d \big( \big \{ \big ( \delta_i, \mathbbm{1}_T (\delta_i) \big ) \big \}_{i \in \bar{m}_d^r} \big)$ is not consistent with $\big \{ \big ( \delta_i, \mathbbm{1}_T (\delta_i) \big ) \big \}_{i \in \bar{I}_r}$.
We thus have

\begin{align}
\mathbb{P}^m &\Big \{ (\delta_1,\ldots,\delta_m) \in \Delta^m :~  d_{\mathbb{P}}(T,H_{\bar{m}_d^r}) > \epsilon \Big \} \nonumber \\
& \leq \mathbb{P}^m \Big \{ \bigcup_{I_r \in \mathcal{I}_r} \big \{  (\delta_1,\ldots,\delta_m) \in \Delta^m :~  H_{m_d^r} \text{ is not consistent with } \big \{ \big ( \delta_i, \mathbbm{1}_T (\delta_i) \big ) \big \}_{i \in I_r} \nonumber \\
&~~~~~~~~~~~~~~~~~~~~~~~~~~~~~~~~~~~~~~~~~~~~~~~~~~~~~~~~~~~~~~~~~~~~~~~~\text{ and } d_{\mathbb{P}}(T,H_{m_d^r}) > \epsilon \big \} \Big \} \label{eq: thm_compr8_relax}\\
& \leq \sum_{I_r \in \mathcal{I}_r} \mathbb{P}^m \Big \{ (\delta_1,\ldots,\delta_m) \in \Delta^m :~  H_{m_d^r} \text{ is not consistent with } \big \{ \big ( \delta_i, \mathbbm{1}_T (\delta_i) \big ) \big \}_{i \in I_r} \nonumber \\
&~~~~~~~~~~~~~~~~~~~~~~~~~~~~~~~~~~~~~~~~~~~~~~~~~~~~~~~~~~~~~~~~~~~~~~~~\text{ and } d_{\mathbb{P}}(T,H_{m_d^r}) > \epsilon \Big \}, \label{eq: thm_compr6_relax} \\
& = {m \choose r} \mathbb{P}^m \Big \{ (\delta_1,\ldots,\delta_m) \in \Delta^m :~  H_{m_d^r} \text{ is not consistent with } \big \{ \big ( \delta_i, \mathbbm{1}_T (\delta_i) \big ) \big \}_{i \in I_r} \nonumber \\
&~~~~~~~~~~~~~~~~~~~~~~~~~~~~~~~~~~~~~~~~~~~~~~~~~~~~~~~~~~~~~~~~~~~~~~~~\text{ and } d_{\mathbb{P}}(T,H_{m_d^r}) > \epsilon \Big \}, \label{eq: thm_compr7_relax}
\end{align}
where \eqref{eq: thm_compr6_relax} is due to the subadditivity of $\mathbb{P}^m$. By \eqref{eq: thm_compr5_relax}, \eqref{eq: thm_compr7_relax} we have that
\begin{align}
\mathbb{P}^m &\Big \{ (\delta_1,\ldots,\delta_m) \in \Delta^m :~  d_{\mathbb{P}}(T,H_{\bar{m}_d^r}) > \epsilon \Big \} \nonumber \\
& \leq {m \choose r} {m-r \choose d} \frac{1}{ {m-d \choose r} } \sum_{i=0}^r {m-d \choose i} \epsilon^i (1-\epsilon)^{m-d-i} \\
& = {m \choose d} \sum_{i=0}^r {m-d \choose i} \epsilon^i (1-\epsilon)^{m-d-i}.
\end{align}

Set $q(m,\epsilon) = {m \choose d} \sum_{i=0}^r {m-d \choose i} \epsilon^i (1-\epsilon)^{m-d-i}$.
Similarly to the last part of the proof of Theorem \ref{thm: prop_viol1_impr}, $\lim_{m \rightarrow \infty} q(m,\epsilon) = 0$.
Construct then algorithm $\big \{ A_m \big \}_{m \geq d+r}$, where $A_m: [\Delta \times \{0,1\}]^m \rightarrow \mathcal{D}$ takes as input a labeled $m$-multisample and returns a hypothesis $H_m = A_m \big( \big \{ \big ( \delta_i, \mathbbm{1}_T (\delta_i) \big ) \big \}_{i=1}^m \big)$ such that $H_m = H_{\bar{m}_d^r}$. By Definition \ref{thm: pac}, algorithm $\big \{ A_m \big \}_{m \geq d+r}$ is PAC-T.
\end{proofof}

\begin{proofof}{Theorem \ref{thm: prop_viol1_impr_relax1}}
The proof of Theorem \ref{thm: prop_viol1_impr_relax1} follows the same lines with the proof of Theorem \ref{thm: prop_viol1_relax} up to equation \eqref{eq: pac_m_r}. Instead of \eqref{eq: pac_m_r},
Theorem \ref{thm: prop_viol1_impr} with $m-r$ in place of $m$ and $\alpha$ in place of $\epsilon$ implies that
\begin{align}
\mathbb{P}^{m-r} \Big \{ \{\delta_i\}_{i \in \{1,\ldots,m\} \setminus I_r} \in \Delta^{m-r} :~ d_{\mathbb{P}}(T,H_{m_d^r}) > \alpha \Big \} = \sum_{i=0}^{d-1} {m \choose i} \epsilon^i (1-\epsilon)^{m-i}. \label{eq: pac_m_r_impr}
\end{align}
By \eqref{eq: errorF_bar}, \eqref{eq: pac_m_r_impr}, we then have that $\bar{F}(\alpha) = 1 - \sum_{i=0}^{d-1} {m \choose i} \alpha^i (1-\alpha)^{m-i}$ for any $\alpha \in [0,1]$.
The last statement together with \eqref{eq: thm_compr1_relax}-\eqref{eq: thm_compr3_relax} leads to

\begin{align}
\mathbb{P}^m &\Big \{ (\delta_1,\ldots,\delta_m) \in \Delta^m :~  H_{m_d^r} \text{ is not consistent with } \big \{ \big ( \delta_i, \mathbbm{1}_T (\delta_i) \big ) \big \}_{i \in I_r} \text{ and } d_{\mathbb{P}}(T,H_{m_d^r}) > \epsilon \Big \} \nonumber \\
& = \int_{\epsilon}^1 \alpha^{r} \bar{F} (\textrm{d}\alpha) \label{eq: thm_compr0_relax_impr} \\
& = \int_{\epsilon}^1 d {m-r \choose d} \alpha^{r} \alpha^{r-1} \big ( 1 - \alpha \big )^{m-r-d} \textrm{d}\alpha \label{eq: thm_compr4_relax_impr} \\
& = \frac{d {m-r \choose d} }{ (r+d) {m \choose r+d} } \sum_{i=0}^{r+d-1} {m \choose i} \alpha^i (1-\alpha)^{m-i},  \label{eq: thm_compr5_relax_impr}
\end{align}
where the equality in \eqref{eq: thm_compr4_relax_impr} (compare with the inequality in \eqref{eq: thm_compr4_relax}) follows from \eqref{eq: thm_compr0_relax_impr} due to the fact that $\bar{F}(\alpha) = 1 - \sum_{i=0}^{d-1} {m \choose i} \alpha^i (1-\alpha)^{m-i}$ and is based on standard integral arguments (see also the proof of Theorem 2.1 in \cite{Garatti}).
Moreover, \eqref{eq: thm_compr5_relax_impr} follows from \eqref{eq: thm_compr4_relax_impr} by repeated integration by parts.

Construct an algorithm as shown above \eqref{eq: thm_compr8_relax} and follow
the same arguments with \eqref{eq: thm_compr8_relax}-\eqref{eq: thm_compr7_relax}. By \eqref{eq: thm_compr5_relax_impr}, \eqref{eq: thm_compr7_relax} we have that
\begin{align}
\mathbb{P}^m &\Big \{ (\delta_1,\ldots,\delta_m) \in \Delta^m :~  d_{\mathbb{P}}(T,H_{\bar{m}_d^r}) > \epsilon \Big \} \nonumber \\
& \leq {m \choose r} \frac{d {m-r \choose d} }{ (r+d) {m \choose r+d} } \sum_{i=0}^{r+d-1} {m \choose i} \alpha^i (1-\alpha)^{m-i} \label{eq: thm_compr9_relax} \\
& = {r +d -1 \choose r} \sum_{i=0}^{r+d-1} {m \choose i} \alpha^i (1-\alpha)^{m-i}. \label{eq: thm_compr10_relax}
\end{align}

Set $q(m,\epsilon) = {r +d -1 \choose r} \sum_{i=0}^{r+d-1} {m \choose i} \alpha^i (1-\alpha)^{m-i}$.
Similarly to the last part of the proof of Theorem \ref{thm: prop_viol1_impr}, $\lim_{m \rightarrow \infty} q(m,\epsilon) = 0$. Construct then algorithm $\big \{ A_m \big \}_{m \geq d+r}$, where $A_m: [\Delta \times \{0,1\}]^m \rightarrow \mathcal{D}$ takes as input a labeled $m$-multisample and returns a hypothesis $H_m = A_m \big( \big \{ \big ( \delta_i, \mathbbm{1}_T (\delta_i) \big ) \big \}_{i=1}^m \big)$ such that $H_m = H_{\bar{m}_d^r}$. By Definition \ref{thm: pac}, algorithm $\big \{ A_m \big \}_{m \geq d+r}$ is PAC-T.
\end{proofof}

\section*{Appendix B: Proofs of Sections \ref{sec: con_opt}}
\begin{proofof}{Theorem \ref{thm: prop_viol}}
Under Assumption \ref{ass: alg}, the hypothesis $H_{m_d} = \big \{ \delta \in \Delta:~ g(x_{d}(\{\delta_i\}_{i \in m_d}),\delta) \leq 0 \big \}$ is consistent with $\big \{ \big ( \delta_i, \mathbbm{1}_T (\delta_i) \big ) \big \}_{i = 1}^m$. This implies that $x_{d}(\{\delta_i\}_{i \in m_d})$ belongs to the feasibility region of $\mathcal{P}[\{\delta_i\}_{i=1}^m]$. Consider an algorithm $\big \{ A_m \big \}_{m \geq d}$, where $A_m: [\Delta \times \{0,1\}]^m \rightarrow \mathcal{D}$ is such that $H_m = A_m \big( \big \{ \big ( \delta_i, \mathbbm{1}_T (\delta_i) \big ) \big \}_{i=1}^m \big)$ with $H_{m} = \big \{ \delta \in \Delta:~ g(x_{m}( \{ \delta_i \}_{i=1}^m),\delta) \leq 0 \big \}$. Moreover, by the theorem hypothesis we have that $x_m(\{ \delta_i \big \}_{i=1}^m) = x_{d}(\{ \delta_i \big \}_{i \in m_d})$, which entails that $H_m = H_{m_d} = G_d \big( \big \{ \big ( \delta_i, \mathbbm{1}_T (\delta_i) \big ) \big \}_{i \in m_d} \big)$, for $G_d$ defined according to \eqref{hyp_G_d}.
Theorem \ref{thm: prop_viol1} implies then that $\big \{ A_{m} \big \}_{m \geq d}$ is PAC-T with $q (m,\epsilon) = {m \choose d} (1-\epsilon)^{m-d}$. The latter, together with the fact that, since $T = \Delta$, $d_{\mathbb{P}}(T,H_m) = \mathbb{P} \big(\{ \delta \in \Delta:~ g(x_m(\{ \delta_i \big \}_{i=1}^m),\delta) > 0 \}\big)$, leads to \eqref{eq: prob_viol}.
\end{proofof}

\begin{proofof}{Proposition \ref{prop: ass_scenApp}}
Fix $d = \zeta$ and consider $m \geq d$.
By the definition of the support constraints, and under Assumption \ref{ass: convex_scenApp}, with $\mathbb{P}^m$-probability one, there exists $m_d( \{ \delta_i \}_{i=1}^m) \in \mathcal{I}_d$ such that $x_m( \{ \delta_i \}_{i=1}^m) = x_{d}( \{ \delta_i \}_{i \in m_d})$ \cite{Campi_Garatti2008}, where $x_m$, $x_{d}$ denote the unique (under Assumption \ref{ass: convex_scenApp}) minimizers of $\mathcal{P}_1[\{\delta_i\}_{i=1}^m]$ and $\mathcal{P}_1[\{\delta_i\}_{i \in m_d}]$, respectively.
The solution $x_{d}( \{ \delta_i \}_{i \in m_d})$ satisfies all constraints that correspond to samples whose indices are not included in $m_d$, otherwise we would not have $x_{d}( \{ \delta_i \}_{i \in m_d}) = x_m$. In other words, $g(x_{d}( \{ \delta_i \}_{i \in m_d}),\delta_i) \leq 0$ for all $i \in \{1,\ldots,m\} \setminus m_d$.
But, since $x_{d}( \{ \delta_i \}_{i \in m_d})$ is the optimal solution of $\mathcal{P}_1[\{\delta_i\}_{i \in m_d}]$ it will satisfy its constraints, i.e. $g(x_{d}( \{ \delta_i \}_{i \in m_d}),\delta_i) \leq 0$ for all $i \in m_d$. Therefore, $g(x_{d}( \{ \delta_i \}_{i \in m_d}),\delta_i) \leq 0 \text{ for all } i \in \{1,\ldots,m\}$ and since $H_{m_d} = \big \{ \delta \in \Delta :~ g(x_{d}( \{ \delta_i \}_{i \in m_d}),\delta) \leq 0 \big \}$, we have that
$\mathbbm{1}_{H_{m_d}}(\delta_i) = 1, \text{ for all } i=1,\ldots,m$.
The last statement together with the fact that $T=\Delta$ implies that the hypothesis $H_{m_d} = G_d \big ( \big \{ \big ( \delta_i, \mathbbm{1}_T (\delta_i) \big ) \big \}_{i \in m_d} \big )$ is consistent with $\big \{ \big ( \delta_i, \mathbbm{1}_T (\delta_i) \big ) \big \}_{i=1}^m$, thus showing that the second part of Assumption \ref{ass: alg} is satisfied.

It remains to show the first part of Assumption \ref{ass: alg}.
For any $I_d \in \mathcal{I}_d$, since $x_{d}( \{ \delta_i \}_{i \in I_d})$ is the minimizer of optimal solution of $\mathcal{P}_1[\{\delta_i\}_{i \in I_d}]$ it will satisfy its constraints, i.e. $g(x_{d}( \{ \delta_i \}_{i \in m_d}),\delta_i) \leq 0$ for all $i \in I_d$. By definition, it then follows that $H_{I_d}$ is consistent with $\big \{ \big ( \delta_i, \mathbbm{1}_T (\delta_i) \big ) \big \}_{i\in I_d}$.
\end{proofof}

\begin{proofof}{Corollary \ref{lm: lemma_scenApp}}
Under Assumption \ref{ass: convex_scenApp}, Proposition \ref{prop: ass_scenApp} shows that $d = \zeta$, $G_d$ satisfy Assumption \ref{ass: alg}. Let then $m_d(\{\delta_i\}_{i=1}^m) \in \mathcal{I}_d$ be a set of indices for which the consistency requirement of Assumption \ref{ass: alg} is satisfied. Moreover, as shown in the proof of Proposition \ref{prop: ass_scenApp}, $x_m(\{ \delta_i \big \}_{i=1}^m) = x_{d}(\{ \delta_i \big \}_{i \in m_d})$. Theorem \ref{thm: prop_viol} leads then to \eqref{eq: prob_viol_scenApp} and concludes the proof.
\end{proofof}

\begin{proofof}{Proposition \ref{prop: fully_sup}}
Assume that $\mathcal{P}_1[\{\delta_i\}_{i=1}^{m}]$ has exactly $d=\zeta \leq m$ support constraints with $\mathbb{P}^m$-probability one.
Under Assumption \ref{ass: convex_scenApp}, and since the number of support constraints is bounded, Proposition \ref{prop: ass_scenApp} implies that Assumption \ref{ass: alg} is satisfied.
However, for the sake of contradiction assume that the second part of Assumption \ref{ass: unique_I} is not satisfied. Therefore, there exist $I_d^1,~I_d^2 \in \mathcal{I}_d$ with $I_d^1 \neq I_d^2$ such that the hypotheses $H_{I_d^1}$ and $H_{I_d^2}$ are both consistent with $\big \{ \big ( \delta_i, \mathbbm{1}_T (\delta_i) \big ) \big \}_{i=1}^{m}$. $H_{I_d^1}$ and $H_{I_d^2}$ are constructed according to \eqref{hyp_G_d} based on the minimizers $x_d( \{ \delta_i\}_{i \in {I_d^1}} \}$ and $x_d( \{ \delta_i\}_{i \in {I_d^2}} \}$ of problems $\mathcal{P}_1[\{\delta_i\}_{i \in I_d^1}]$ and $\mathcal{P}_1[\{\delta_i\}_{i \in I_d^2}]$, respectively. By the definition of consistency and the form of $H_{I_d^1}$, $H_{I_d^2}$, we have that $x_d( \{ \delta_i\}_{i \in {I_d^1}} \}$ and $x_d( \{ \delta_i\}_{i \in {I_d^2}} \}$
satisfy also all constraints corresponding to $\{\delta_i\}_{i \in \{1,\dots,m\} \setminus I_d^1}$ and $\{\delta_i\}_{i \in \{1,\dots,m\} \setminus I_d^1}$, respectively.
Therefore, and under the uniqueness part of Assumption \ref{ass: convex_scenApp}, $x_d( \{ \delta_i\}_{i \in {I_d^1}} \} = x_d( \{ \delta_i\}_{i \in {I_d^2}} \} = x_m(\{\delta_i\}_{i=1}^m)$, where $x_m$ is the  minimizer of $\mathcal{P}_1[\{\delta_i\}_{i=1}^{m}]$.

By the definition of the support constraints (see Definition 4 in \cite{Calafiore_Campi2006} and discussion in Section \ref{sec: scenApp}) and under the assumption that $\mathcal{P}_1[\{\delta_i\}_{i=1}^{m}]$ has exactly $d$ support constraints, the last statement implies that the constraints that correspond to samples with indices in $I_d^1$ and $I_d^2$ are support constraints. Moreover, the fact that $I_d^1 \neq I_d^2$ would imply that there exist at least one index that does not belong to both $I_d^1$ and $I_d^2$. Since $|I_d^1|=|I_d^2|=d$, the last statement implies that the number of support constraints would be greater than or equal to $d+1$, thus contradicting the assumption that we have exactly $d$ support constraints, proving the second part of Assumption \ref{ass: unique_I}.
To conclude the proof it remains to show the first part of Assumption \ref{ass: unique_I}; this is the same with the first statement of Assumption \ref{ass: alg} and can be shown as in the proof of Proposition \ref{prop: ass_scenApp}.
\end{proofof}

\begin{proofof}{Corollary \ref{lm: lemma_scenApp_impr}}
Under Assumption \ref{ass: convex_scenApp}, and since $\mathcal{P}_1[\{\delta_i\}_{i=1}^{m}]$ has $d = \zeta \leq m$ support constraints with $\mathbb{P}^m$-probability one, Proposition \ref{prop: fully_sup} shows that $d$, $G_d$ satisfy Assumption \ref{ass: unique_I}.
Similarly to the proof of Corollary \ref{lm: lemma_scenApp}, let $m_d(\{\delta_i\}_{i=1}^m) \in \mathcal{I}_d$ be the unique set of indices for which the consistency requirement of Assumption \ref{ass: unique_I} is satisfied. Moreover, as shown in the proof of Proposition \ref{prop: ass_scenApp}, $x_m(\{ \delta_i \big \}_{i=1}^m) = x_{d}(\{ \delta_i \big \}_{i \in m_d})$. Theorem \ref{thm: prop_viol} leads then to \eqref{eq: prob_viol_scenApp_impr} and concludes the proof.
\end{proofof}

\begin{proofof}{Proposition \ref{prop: ass_boxApp}}
We first show that with $\mathbb{P}^m$-probability one, there exists a unique set $m_d(\{\delta_i\}_{i=1}^{m}) \in \mathcal{I}_d$ with $d=2n_\delta$ such that $B(p_{d}(\{\delta_i\}_{i \in m_d})) = B(p_m(\{\delta\}_{i=1}^m))$.
$B(p_{d}(\{\delta_i\}_{i \in m_d}))$ is the minimum volume hyper-rectangle that contains $d=2n_\delta$ samples of the uncertainty with indices given by $m_d$.
Clearly, $B(p_{d}(\{\delta_i\}_{i \in I_d})) \subseteq B(p_m)$ for any $I_d \in \mathcal{I}_d$. Therefore, it suffices to show that there exists $I_d \in \mathcal{I}_d$ such that $B(p_m) = B(p_{d}(\{\delta_i\}_{i \in m_d}))$.
Let $B(p_m) = \times_{\ell=1}^{n_\delta} \big [ \underline{p}_{m}^{\ell},~ \overline{p}_{m}^{\ell} \big ]$, where $\underline{p}_{m}^{\ell},~ \overline{p}_{m}^{\ell}$ denote the $\ell$-th elements of
$\underline{p}_{m}$ and $\overline{p}_{m}$, respectively. By inspection of $\widetilde{\mathcal{P}}_2[\{\delta_i\}_{i=1}^{m}]$, for all $\ell=1,\ldots,n_\delta$,
$\underline{p}_m^{\ell} = \min_{i=1,\ldots,m} \delta_i^\ell$ and $\overline{p}_{m}^{*,\ell} = \max_{i=1,\ldots,m} \delta_i^\ell$, where $\delta_i^\ell$ denotes the $\ell$-th element of sample $i$.

With $\mathbb{P}^m$-probability one $\widetilde{\mathcal{P}}_2[\{\delta_i\}_{i=1}^{m}]$ admits a unique solution. Let $\underline{i}^\ell = \arg \min_{i=1,\ldots,m} \delta_i^\ell$ and $\overline{i}^\ell = \arg \max_{i=1,\ldots,m} \delta_i^\ell$, for $\ell=1,\ldots,n_\delta$. Consider then the set of indices $m_d = \big \{ \{\underline{i}^\ell, \overline{i}^\ell\}_{\ell=1}^{n_\delta} \big \}$. With $\mathbb{P}^m$-probability one, $m_d$ is unique, $|m_d| = 2n_\delta$ and by construction (see the definition of $B(p_m)$) $B(p_{d}(\{\delta_i\}_{i \in m_d})) = B(p_m)$.

Fix $d = 2n_\delta$ and consider $m \geq d$. Let $m_d(\{\delta_i\}_{i=1}^m) \in \mathcal{I}_d$ be the unique set of indices for which $B(p_{d}(\{\delta_i\}_{i \in m_d})) = B(p_m)$ and denote by $\mathcal{X}_{m_d} = \{x \in \mathcal{X} :~ g(x,\delta) \leq 0,\, \forall \delta \in B(p_{d}(\{\delta_i\}_{i \in m_d})) \}$ the feasibility region of $\mathcal{P}_2[\{\delta_i\}_{i \in m_d}]$, which is non-empty by Assumption \ref{ass: boxApp}. Let then
$x_{d}(\{\delta_i\}_{i \in m_d})$ be a minimizer of $\mathcal{P}_2[\{\delta_i\}_{i \in m_d}]$.
By construction $x_{d}(\{\delta_i\}_{i \in m_d}) \in \mathcal{X}_{m_d}$, so it would satisfy all constraints of $\mathcal{P}_2[\{\delta_i\}_{i \in m_d}]$. Therefore, $g(x_{d}(\{\delta_i\}_{i \in m_d}),\delta) \leq 0$ for all $\delta \in B(p_{d}(\{\delta_i\}_{i \in m_d}))$. Since $B(p_{d}(\{\delta_i\}_{i \in m_d})) = B(p_m)$, the last statement is equivalent to
\begin{align}
    g(x_{d}(\{\delta_i\}_{i \in m_d}),\delta) \leq 0 \text{ for all } \delta \in B(p_{m}). \label{eq: feas_x_I_box}
\end{align}
The hypothesis $H_{m_d}$ is given by $H_{m_d} = \big \{ \delta \in \Delta :~ g(x_{d}(\{\delta_i\}_{i \in m_d}),\delta) \leq 0 \big \}$. By \eqref{eq: feas_x_I_box}, this implies that
\begin{align}
    \mathbbm{1}_{H_{m_d}}(\delta) = 1, \text{ for all } \delta \in B(p_{m}).
\end{align}
Since $B(p_{m})$ contains all samples $\delta_1,\ldots,\delta_m$, the last statement implies that
$\mathbbm{1}_{H_{m_d}}(\delta_i) = 1$ for all $i=1,\ldots,m$.
The last statement together with the fact that $T=\Delta$ implies that the hypothesis $H_{m_d} = G_d \big ( \big \{ \big ( \delta_i, \mathbbm{1}_T (\delta_i) \big ) \big \}_{i \in m_d} \big )$ is consistent with $\big \{ \big ( \delta_i, \mathbbm{1}_T (\delta_i) \big ) \big \}_{i=1}^m$, thus showing that the second part of Assumption \ref{ass: alg} is satisfied.
To conclude the proof it remains to show the first part of Assumption \ref{ass: alg}; this can be done as in the proof of Proposition \ref{prop: ass_scenApp}.
\end{proofof}

\begin{proofof}{Corollary \ref{lm: lemma_boxApp}}
Under Assumption \ref{ass: boxApp}, Proposition \ref{prop: ass_boxApp} shows that $d = 2n_\delta$, $G_d$ satisfy Assumption \ref{ass: alg}. Let then $m_d(\{\delta_i\}_{i=1}^m) \in \mathcal{I}_d$ be the unique (under Proposition \ref{prop: ass_boxApp}) set of indices for which the consistency requirement of Assumption \ref{ass: alg} is satisfied. Moreover, as shown in the proof of Proposition \ref{prop: ass_scenApp}, $B(p_m(\{ \delta_i \big \}_{i=1}^m)) = B(p_{d}(\{ \delta_i \big \}_{i \in m_d}))$, which implies that $\mathcal{X}_{m_d} = \mathcal{X}_m$. Due to the uniqueness part of Assumption \ref{ass: boxApp} we then have that $x_{d}( \{ \delta_i \}_{i \in m_d}) = x_m( \{ \delta_i \}_{i=1}^m )$.
Theorem \ref{thm: prop_viol} leads then to \eqref{eq: prob_viol_boxApp} and concludes the proof.
\end{proofof}

\begin{proofof}{Proposition \ref{prop: suff_box}}
If \eqref{eq: suff_box} is satisfied, then, with $\mathbb{P}^m$-probability one, for all $I_d \in \mathcal{I}_d$
\begin{align}
\big \{ \delta \in \Delta :~ g(x_{d}(\{\delta_i\}_{i \in I_d}),\delta) > 0 \big \} =  \big \{ \delta \in \Delta :~ \delta \notin B(p_{d}(\{\delta_i\}_{i \in I_d})) \big \}. \label{eq: suff1}
\end{align}
As shown in the proof of Proposition \ref{prop: ass_boxApp}, there exists a unique $m_d \in \mathcal{I}_d$ such that $B(p_{d}(\{\delta_i\}_{i \in m_d})) = B(p_m(\{\delta_i\}_{i = 1}^m))$. For any $I_d \in \mathcal{I}_d$ with $I_d \neq m_d$ we have that $B(p_{d}(\{\delta_i\}_{i \in I_d})) \subset B(p_m(\{\delta_i\}_{i = 1}^m)) = B(p_{d}(\{\delta_i\}_{i \in m_d}))$.

For the sake of contradiction assume that second part of Assumption \ref{ass: unique_I} is not satisfied. This implies that there exists $I_d \in \mathcal{I}_d$ with $I_d \neq m_d$ such that $H_{I_d}$ is consistent with $\big \{ \big ( \delta_i, \mathbbm{1}_T (\delta_i) \big ) \big \}_{i=1}^{m}$.
Since $H_{I_d} = \{\delta \in \Delta :~ g(x_{d}(\{\delta_i\}_{i \in I_d}),\delta) \leq 0 \}$ ($x_{d}$ is a minimizer of $\mathcal{P}_2[\{\delta_i\}_{i \in I_d}]$), consistency implies that $g(x_{d}(\{\delta_i\}_{i \in I_d})),\delta_i) \leq 0$ for all $i=1,\ldots,m$. Since $B(p_{d}(\{\delta_i\}_{i \in I_d})) \subset B(p_m)$, the last statement implies that there exists $\ell \in \{1,\ldots,m\} \setminus I_d$ such that $\delta_\ell \notin B(p_{d}(\{\delta_i\}_{i \in I_d}))$ and $g(x_{d}(\{\delta_i\}_{i \in I_d}),\delta_\ell) \leq 0$, i.e. there exists at least one uncertainty realization $\delta$ that is not contained in $B(p_{d}(\{\delta_i\}_{i \in I_d}))$ and does not lead to constraint violation. Therefore,
\begin{align}
\big \{ \delta \in \Delta :~ g(x_{d}(\{\delta_i\}_{i \in I_d}),\delta) > 0 \big \} \subset \big \{ \delta \in \Delta :~ \delta \notin B(p_{d}(\{\delta_i\}_{i \in I_d})) \big \}. \label{eq: suff2}
\end{align}
Equations \eqref{eq: suff1} and \eqref{eq: suff2} establish a contradiction, proving the second part of Assumption \ref{ass: unique_I}.
To conclude the proof it remains to show the first part of Assumption \ref{ass: unique_I}; this is the same with the first statement of Assumption \ref{ass: alg} and can be shown as in the proof of Proposition \ref{prop: ass_scenApp}.
\end{proofof}

\begin{proofof}{Corollary \ref{lm: lemma_boxApp_impr}}
Under Assumption \ref{ass: boxApp}, and since \eqref{eq: suff_box} is satisfied with $\mathbb{P}^m$-probability one, by Proposition \ref{prop: suff_box} we have that Assumption \ref{ass: unique_I} is satisfied.
Equation \eqref{eq: prob_viol_boxApp_impr} results then from \eqref{eq: pac_tight} in Theorem \ref{thm: prop_viol1_impr}, following similar arguments with the proof of Corollary \ref{lm: lemma_boxApp}.
\end{proofof}

\section*{Appendix C: Proofs of Sections \ref{sec: seq_opt}}
\begin{proofof}{Proposition \ref{prop: cons_cascade}}
Under Assumption \ref{ass: convex_scenApp}, Assumption \ref{ass: alg} is satisfied for $d_1 \in \mathbb{N}$, $G_{d_1}: [\Delta \times \{0,1\}]^{d_1} \rightarrow \mathcal{D}$. Then, there exists $m_{d_1}(\{\delta_i\}_{i=1}^m) \in \mathcal{I}_{d_1}$ such that $H_{m_{d_1}} =  G_{d_1} \big( \big \{ \big ( \delta_i, \mathbbm{1}_T (\delta_i) \big ) \big \}_{i \in m_{d_1}} \big )$ is consistent with
$\big \{ \big ( \delta_i, \mathbbm{1}_T (\delta_i) \big ) \big \}_{i=1}^{m}$. Since $H_{m_{d_1}} = \big \{\delta \in \Delta :~ g \big ( x_{d_1}(\{\delta_i\}_{i \in m_{d_1}}), \delta \big ) \leq 0 \big \}$,
\begin{align}
g \big ( x_{d_1}(\{\delta_i\}_{i \in m_{d_1}}), \delta_i \big ) \leq 0, \text{ for all } i=1,\ldots,m. \label{eq: cascade1}
\end{align}

Moreover, under Assumption \ref{ass: convex_scenApp}, for all $x \in \mathcal{X}$, Assumption \ref{ass: alg} is satisfied for $d_2 \in \mathbb{N}$, $\widetilde{G}_{d_2}[x]: [\Delta \times \{0,1\}]^{d_2} \rightarrow \mathcal{D}$. This implies that,
for all $x \in \mathcal{X}$, there exists $m_{d_2}[x](\{\delta_i\}_{i=1}^m) \in \mathcal{I}_{d_2}$ such that the hypothesis $\widetilde{H}_{m_{d_2}[x]}[x] =  \widetilde{G}_{d_2}[x] \big( \big \{ \big ( \delta_i, \mathbbm{1}_T (\delta_i) \big ) \big \}_{i \in m_{d_2}[x]} \big )$ is consistent with $\big \{ \big ( \delta_i, \mathbbm{1}_T (\delta_i) \big ) \big \}_{i=1}^{m}$.
Since $\widetilde{H}_{m_{d_2}[x]}[x] = \big \{ \delta \in \Delta :~ \widetilde{g} \big ( y_{d_2}[x](\{\delta_i\}_{i \in m_{d_2}[x]},x,\delta)\big ) \leq 0 \big \}$, for any $x \in \mathcal{X}$,
\begin{align}
\widetilde{g} \big ( y_{d_2}[x](\{\delta_i\}_{i \in m_{d_2}[x]},x,\delta_i)\big ) \leq 0, \text{ for all } i=1,\ldots,m. \label{eq: cascade2}
\end{align}

Set $d = d_1 + d_2$ and consider $m \geq d$. Choose $m_d(\{\delta_i\}_{i=1}^m) \in \mathcal{I}_d$ such that $m_d(\{\delta_i\}_{i=1}^m) \supseteq m_{d_1}(\{\delta_i\}_{i=1}^m) \cup m_{d_2}[x_{d_1}(\{\delta_i\}_{i \in m_{d_1}(\{\delta_i\}_{i=1}^m)})](\{\delta_i\}_{i=1}^m)$ (we do not have equality since some indices may belong to both $m_{d_1}$ and $m_{d_2}[x]$, implying that some constraints are of support for both problems in the cascade), where $x_{d_1}(\{\delta_i\}_{i \in m_{d_1}})$ is the minimizer of $\mathcal{P}[\{\delta_i\}_{i \in m_{d_1}}]$ that is used to construct $H_{m_{d_1}}$. For simplicity, as in \eqref{eq: cascade1}, \eqref{eq: cascade2}, we do not show the argument $(\{\delta_i\}_{i=1}^m)$ of $m_{d_1}$, $m_{d_2}[x_{d_1}(\{\delta_i\}_{i \in m_{d_1}})]$. As shown in the proof of Proposition \ref{prop: ass_scenApp}, since $m_d \supseteq m_{d_1}$, $x_d(\{\delta_i\}_{i \in m_d}) = x_{d_1}(\{\delta_i\}_{i \in m_{d_1}})$. Therefore, \eqref{eq: cascade1} implies that $g \big ( x_{d}(\{\delta_i\}_{i \in m_{d}}), \delta_i \big ) \leq 0, \text{ for all } i=1,\ldots,m$.
We also have that $m_d \supseteq m_{d_2}[x_{d_1}(\{\delta_i\}_{i \in m_{d_1}})] = m_{d_2}[x_{d}(\{\delta_i\}_{i \in m_{d}})]$,
where the last equality follows from the fact that $x_d(\{\delta_i\}_{i \in m_d}) = x_{d_1}(\{\delta_i\}_{i \in m_{d_1}})$.
Similarly to the previous case, as shown in the proof of Proposition \ref{prop: ass_scenApp} we have that $y_d[x_d(\{\delta_i\}_{i \in m_d})] (\{\delta_i\}_{i \in m_d}) = y_{d_2}[x_d(\{\delta_i\}_{i \in m_d})] (\{\delta_i\}_{i \in m_{d_2}[x_d(\{\delta_i\}_{i \in m_d})]})$.
By \eqref{eq: cascade2}, $\widetilde{g} \big ( y_d[x_d(\{\delta_i\}_{i \in m_d})] (\{\delta_i\}_{i \in m_d}),x_d(\{\delta_i\}_{i \in m_d}),\delta_i)\big ) \leq 0, \text{ for all } i=1,\ldots,m$.
Therefore, we have that
\begin{align}
g \big ( x_{d}(\{\delta_i\}_{i \in m_{d}}), & \delta_i \big ) \leq 0
\text{ and } \nonumber \\ &\widetilde{g} \big ( y_d[x_d(\{\delta_i\}_{i \in m_d})] (\{\delta_i\}_{i \in m_d}),x_d(\{\delta_i\}_{i \in m_d}),\delta_i)\big ) \leq 0, \text{ for all } i=1,\ldots,m. \label{eq: cascade3}
\end{align}
Since $T = \Delta$, \eqref{eq: cascade3}, \eqref{eq:cascade_end} imply that
$G_d^c \big( \big \{ \big ( \delta_i, \mathbbm{1}_T (\delta_i) \big ) \big \}_{i \in m_d} \big) = H_{m_{d}} \cap \widetilde{H}_{m_{d}}[x_d(\{\delta_i\}_{i \in m_{d}})]$ is consistent with $\big \{ \big ( \delta_i, \mathbbm{1}_T (\delta_i) \big ) \big \}_{i=1}^{m}$. To conclude the proof it remains to show the first part of Assumption \ref{ass: alg}; this can be done as in the proof of Proposition \ref{prop: ass_scenApp}.
\end{proofof}

\begin{proofof}{Theorem \ref{thm: cons_cascade_opt}}
Under Assumption \ref{ass: convex_scenApp}, Proposition \ref{prop: cons_cascade} implies that
$G_d^c$ satisfies Assumption \ref{ass: alg}. Then, there exists $m_d \in \mathcal{I}_d$  such that the hypothesis
$H_{m_d}^c = \big \{ \delta \in \Delta :~  \big ( g(x_{m_d},\delta) \leq 0 \big ) \text{ and } \big ( \widetilde{g}(y_{m_d}[x_{m_d}],x_{m_d},\delta) \leq 0 \big ) \big \}$ is consistent with $\big \{ \big ( \delta_i, \mathbbm{1}_T (\delta_i) \big ) \big \}_{i=1}^{m}$. Consider an algorithm $\big \{ A_m \big \}_{m \geq d}$, where $A_m: [\Delta \times \{0,1\}]^m \rightarrow \mathcal{D}$ is such that $H_m = A_m \big( \big \{ \big ( \delta_i, \mathbbm{1}_T (\delta_i) \big ) \big \}_{i=1}^m \big)$ with $H_m = \big \{ \delta \in \Delta :~  \big ( g(x_{m},\delta) \leq 0 \big ) \text{ and } \big ( \widetilde{g}(y_{m}[x_{m}],x_{m},\delta) \leq 0 \big ) \big \}$. Under Assumption \ref{ass: convex_scenApp}, following the proof of Proposition \ref{prop: ass_scenApp} we have that $x_m(\{ \delta_i \big \}_{i=1}^m) = x_{d}(\{ \delta_i \big \}_{i \in m_d})$ and $y_m[x_m](\{ \delta_i \big \}_{i=1}^m) = y_{d}[x_d](\{ \delta_i \big \}_{i \in m_d})$ and hence $H_m = H_{m_d}^c = G_d^c \big( \big \{ \big ( \delta_i, \mathbbm{1}_T (\delta_i) \big ) \big \}_{i \in m_d} \big)$. Theorem \ref{thm: prop_viol1} implies then that $\big \{ A_{m} \big \}_{m \geq d}$ is PAC-T with $q (m,\epsilon) = {m \choose d} (1-\epsilon)^{m-d}$. The latter, together with the fact that, since $T = \Delta$, $d_{\mathbb{P}} (T,H_{m}) = \mathbb{P} \Big ( \delta \in \Delta :~ \big ( g(x_{m},\delta) > 0 \big ) \text{ or } \big ( \widetilde{g}(y_{m}[x_{m}],x_{m},\delta) > 0 \big ) \Big )$, leads to \eqref{eq: pac_cascade_opt}.
\end{proofof}

%\footnotesize{
%\bibliography{IEEEabrv,biblio,scApproach_compression_ref}

\begin{thebibliography}{10}
\providecommand{\url}[1]{#1}
\csname url@samestyle\endcsname
\providecommand{\newblock}{\relax}
\providecommand{\bibinfo}[2]{#2}
\providecommand{\BIBentrySTDinterwordspacing}{\spaceskip=0pt\relax}
\providecommand{\BIBentryALTinterwordstretchfactor}{4}
\providecommand{\BIBentryALTinterwordspacing}{\spaceskip=\fontdimen2\font plus
\BIBentryALTinterwordstretchfactor\fontdimen3\font minus
  \fontdimen4\font\relax}
\providecommand{\BIBforeignlanguage}[2]{{%
\expandafter\ifx\csname l@#1\endcsname\relax
\typeout{** WARNING: IEEEtran.bst: No hyphenation pattern has been}%
\typeout{** loaded for the language `#1'. Using the pattern for}%
\typeout{** the default language instead.}%
\else
\language=\csname l@#1\endcsname
\fi
#2}}
\providecommand{\BIBdecl}{\relax}
\BIBdecl

\bibitem{Floyd_Warmuth1995}
S.~Floyd and M.~Warmuth, ``{Sample compression, learnability, and the
  Vapnik-Chervonenkis dimension},'' \emph{Machine Learning}, pp. 1--36, 1995.

\bibitem{Robust_Opt_book}
A.~Ben-Tal, L.~El-Ghaoui, and A.~Nemirovski, \emph{{Robust
  Optimization}}.\hskip 1em plus 0.5em minus 0.4em\relax Princeton Series in
  Applied Mathematics, 2009.

\bibitem{prekopa}
A.~Prekopa, \emph{{Stochastic Programming}}.\hskip 1em plus 0.5em minus
  0.4em\relax Cluwer Academic Publishers, Dordrecht, Boston, 1995.

\bibitem{Shapiro}
A.~Shapiro, ``{Stochastic programming approach to optimization under
  uncertainty },'' \emph{Mathematical Programming, Series B}, vol. 112, pp. 183
  -- 183, 2008.

\bibitem{NemShap}
A.~Nemirovski and A.~Shapiro, ``{Convex Approximations of Chance Constrained
  Programs},'' \emph{Siam Journal on Control and Optimization}, vol.~17, no.~4,
  pp. 969 -- 996, 2006.

\bibitem{Berts2}
D.~Bertsimas and M.~Sim, ``{Tractable Approximations to Robust Conic
  Optimization Problems},'' \emph{Mathematical Programming, Series B}, vol.
  107, pp. 5--36, 2006.

\bibitem{Calafiore_Campi2006}
G.~Calafiore and M.~Campi, ``The scenario approach to robust control design,''
  \emph{IEEE Transactions on Automatic Control}, vol.~51, no.~5, pp. 742--753,
  2006.

\bibitem{Campi_Garatti2008}
M.~Campi and S.~Garatti, ``The exact feasibility of randomized solutions of
  uncertain convex programs,'' \emph{SIAM Journal on Optimization}, vol.~19,
  no.~3, pp. 1211--1230, 2008.

\bibitem{Calafiore}
G.~Calafiore, ``{Random Convex Programs},'' \emph{SIAM Journal on
  Optimization}, vol.~20, no.~6, pp. 3427--3464, 2010.

\bibitem{VCtheory}
V.~Vapnik and A.~Chervonenkis, ``{On the uniform convergence of relative
  frequencies of events to their probabilities},'' \emph{Theory Probab. Appl.},
  vol.~16, no.~2, pp. 264 -- 280, 1971.

\bibitem{Vapnik}
V.~Vapnik, \emph{{Statistical Learning Theory}}.\hskip 1em plus 0.5em minus
  0.4em\relax John Wiley \& Sons, Inc., 1998.

\bibitem{Anthony_Biggs}
M.~Anthony and N.~Biggs, \emph{{Computational Learning Theory}}.\hskip 1em plus
  0.5em minus 0.4em\relax Cambridge Tracts in Theoretical Computer Science,
  1992.

\bibitem{Vidyasagar1997}
M.~Vidyasagar, \emph{{A Theory of Learning and Generalization}}.\hskip 1em plus
  0.5em minus 0.4em\relax London, U.K.: Springer-Verlag, 1997.

\bibitem{book_Tempo}
R.~Tempo, G.~Calafiore, and F.~Dabbene, \emph{{Randomized Algorithms for
  Analysis and Control of Uncertain Systems}}.\hskip 1em plus 0.5em minus
  0.4em\relax Springer-Verlag, London, 2005.

\bibitem{AlamoVC}
T.~Alamo, R.~Tempo, and E.~Camacho, ``{Randomized strategies for probabilistic
  solutions of uncertain feasibility and optimization problems},'' \emph{IEEE
  Transactions on Automatic Control}, vol.~54, no.~11, pp. 2545 -- 2559, 2009.

\bibitem{Margellos2013}
\BIBentryALTinterwordspacing
K.~Margellos, P.~Goulart, and J.~Lygeros, ``On the road between robust
  optimization and the scenario approach for chance constrained optimization
  problems,'' \emph{IEEE Transactions on Automatic Control, to appear}, 2014.
  [Online]. Available:
  \url{http://control.ee.ethz.ch/index.cgi?page=publications&action=details&id=4259}
\BIBentrySTDinterwordspacing

\bibitem{Grammatico_2013}
\BIBentryALTinterwordspacing
S.~Grammatico, X.~Zhang, K.~Margellos, P.~Goulart, and J.~Lygeros, ``{A
  scenario approach to non-convex control design},'' \emph{Technical Report,
  ETH Z\"urich}, 2013. [Online]. Available:
  \url{http://control.ee.ethz.ch/~gsergio/GraZhaMarGouLyg_TAC13.pdf}
\BIBentrySTDinterwordspacing

\bibitem{Garatti}
M.~Campi and S.~Garatti, ``A sampling-and-discarding approach to
  chance-constrained optimization: feasibility and optimality,'' \emph{Journal
  of Optimization Theory and Applications}, vol. 148, no.~2, pp. 257--280,
  2011.

\bibitem{Schildbach_2013}
G.~Schildbach, L.~Fagiano, and M.~Morari, ``{Randomized Solutions to Convex
  Programs with Multiple Chance Constraints},'' \emph{SIAM Journal on
  Optimization}, vol.~23, no.~4, pp. 2479 -- 2501, 2013.

\bibitem{Deori_2013}
L.~Deori, S.~Garatti, and M.~Prandini, ``{Stochastic constrained control:
  trading performance for state constraint feasibility},'' \emph{Proceeding of
  European Control Conference}, pp. 2740--2745, 2013.

\bibitem{Alamo}
T.~Alamo, R.~Tempo, and A.~Luque, ``{On the Sample Complexity of Randomized
  Approaches to the Analysis and Design under Uncertainty},'' \emph{American
  Control Conference}, pp. 4671 -- 4676, 2010.

\end{thebibliography}
%\bibliographystyle{IEEEtran}
%} \normalsize

% Generated by IEEEtran.bst, version: 1.13 (2008/09/30)

\end{document}